**Towards Enhancing Data Equity in Public Health Data Science**

Yiran Wang, PhD[1], Alicia E Boyd, PhD[1], Lillian Rountree, MS[1], Yi Ren, MS[1], Kate Nyhan, MLS[2,3], Ruchit Nagar, MD[4,5], Jackson Higginbottom, MPH[1], Megan L Ranney, MD[6], Harsh Parikh, PhD[1,7], Bhramar Mukherjee, PhD[1,8]

[1] Department of Biostatistics, Yale School of Public Health, Yale University, 60 College Street, New Haven, CT 06510, USA
[2] Harvey Cushing/John Hay Whitney Medical Library, Yale University, 333 Cedar St, New Haven, CT 06510, USA
[3] Department of Environmental Health Sciences, Yale School of Public Health, Yale University, 60 College Street, New Haven, CT 06510, USA
[4] Department of Pediatrics, Yale – New Haven Hospital, New Haven, CT 06510, USA
[5] Department of Internal Medicine, Yale – New Haven Hospital, New Haven, CT 06410, USA
[6] Department of Health Policy and Management, Yale School of Public Health, Yale University, 60 College Street, New Haven, CT 06510, USA
7 Department of Biostatistics, Johns Hopkins Bloomberg School of Public Health, Baltimore, MD 21205, USA
[8] Department of Chronic Disease Epidemiology, Yale University, New Haven, CT 06510, USA

Corresponding author:
Bhramar Mukherjee
60 College St
New Haven, CT 06510
bhramar.mukerhjee@yale.edu
(203) 737-8644

# Contents

**ABSTRACT**

In public health, data-driven decisions profoundly influence policies, interventions, and prevention strategies. However, acute disparities in data representation across populations persist, often leading to skewed insights and suboptimal decisions. Recognizing, quantifying, and addressing these challenges require a structured roadmap that integrates insights across domains — including, but not limited to, public health data science and computer science — and critically examines these insights through reflexivity and critical theory. This need has brought increasing attention to the concept of data equity, which offers a guiding framework for addressing systemic bias in data use. Data equity aims to ensure the fair and inclusive representation, collection, and use of data to prevent the introduction or exacerbation of systemic biases that could lead to invalid downstream inference and decisions. We highlight the urgency of this issue by presenting three public health examples where the acute lack of representative datasets and skewed knowledge adversely affect decision-making across diverse sub-groups. The challenges illustrated in these examples mirror broader concerns raised in both public health and computer science literature. While existing public health literature emphasizes the paucity of high-quality data from specific sub-populations, computer science and statistical literature offer general criteria and metrics for assessing biases in data and modeling systems. Building upon foundational concepts from these fields, we propose a working definition of public health data equity and introduce a structured framework for self-auditing public health data science practices. This framework integrates core principles from computational science, such as fairness, accountability, transparency, ethics, privacy, and confidentiality, with key public health considerations, including selection bias, representativeness, generalizability, causality, and information bias. Our framework aims to guide public health researchers in evaluating and improving equity throughout the entire data life cycle: from design and collection, to measurement, analysis, interpretation, and translation. By fundamentally embedding data equity within public health research and practice, this work provides a multidisciplinary pathway toward ensuring that data-driven policies, artificial intelligence innovations, and emerging technologies foster improved health outcomes and well-being for all populations. We conclude by emphasizing the critical understanding that, although data equity is an essential first step, it does not inherently guarantee information, learning, or decision equity.

1. Introduction

As data-driven technologies continue to shape decision-making for the public, concerns about ethics and equity in data practices have been surfacing to increased priority and prominence (Price and Cohen, 2019, Carroll et al., 2020, Wesson et al., 2022, Ruijer et al., 2023). Data equity, an evolving concept, emphasizes fairness, accessibility, and impartiality in the use and distribution of data to promote human rights, dignity, and opportunity for all communities (World Economic Forum, 2024). At its core, data equity recognizes that the ways in which data are collected, processed, and applied have profound consequences on social, economic, learning, and health outcomes. However, many current data systems perpetuate distorted inference. This can often be due to the lack of representation in datasets, the use of modeling practices that overlook structural disparities, and implicit biases ingrained in algorithmic decision-making (Ntoutsi et al., 2020, Ferrara, 2023, Ray, 2023, Hadi et al., 2024). These issues are particularly concerning in the realm of public health, where data inform decisions, policies, interventions, quality and delivery of healthcare at an individual, community and national level — impacting population health at different scales.

**Public Health Challenge.** Public health is inherently an interdisciplinary field, integrating expertise from epidemiology, biostatistics, health policy, and social sciences to address complex health challenges. Public health experts are involved in every stage of the health data cycle, from study design and data collection to statistical modeling, inference, and policy evaluation (Turnock, 2016). A central tenet of public health is striving to achieve best possible health for everyone regardless of their demographics, socio-economics, and geography, referred to as health equity (Kawachi, 2002, Braveman, 2003). Achieving health equity requires access to high-quality representative data, as well as analytical frameworks that account for differences in data availability, representation, and usage. Without such considerations, data-driven policies and interventions risk reinforcing inequities rather than addressing them.

Historically, health disparities have been exacerbated by implicit or explicit bias in data collection practices, exclusionary study designs, and limited consideration of marginalized communities in health research. These gaps not only affect scientific validity but also lead to real-world consequences, such as biased clinical algorithms, inadequate resource allocation, and ineffective public health interventions. As conversational AI tools and large multi-model models becomes increasingly accessible, public health researchers need clear guidelines on how to quantify, evaluate, and implement data equity principles in real-world context (Subramonian et al., 2024, Oketch et al., 2025).

**Contributions.** This paper unites the theoretical and methodological foundations of data equity in computer science and public health data science to propose a structured framework for equitable public health data practices:

- We begin by examining three case studies from public health that illustrate how the lack of data equity hinders scientific progress and leads to sub-optimal health decisions.
- Next, we present contemporary conceptualization and operationalization of data equity in both public health as well as computer science:
    - Contemporary research in computer science has focused on fairness, accountability, transparency, ethics, privacy, and confidentiality to provide

guidelines for mitigating (algorithmic) bias and ensuring responsible data governance.

     o Similarly, public health data science — rooted in biostatistics and epidemiology — also emphasizes careful consideration of selection bias, representativeness, generalizability, causality, and information bias as key considerations in producing reliable and equitable health insights.

- We build on these principles and propose a framework to integrate data equity into the public health data science workflow (see Figure 1). Our approach is rooted in the above-mentioned principles as well as asking reflexive questions that emerge from domain science.

- Finally, we propose a *self-auditing* implementable process across the data life cycle to embrace and enhance data equity in the practice of public health data science. We end with a sobering conclusion that data equity is just an initial and necessary step but does not equate or imply learning/predictive/information equity. Despite achieving data equity, decision equity may remain distant.

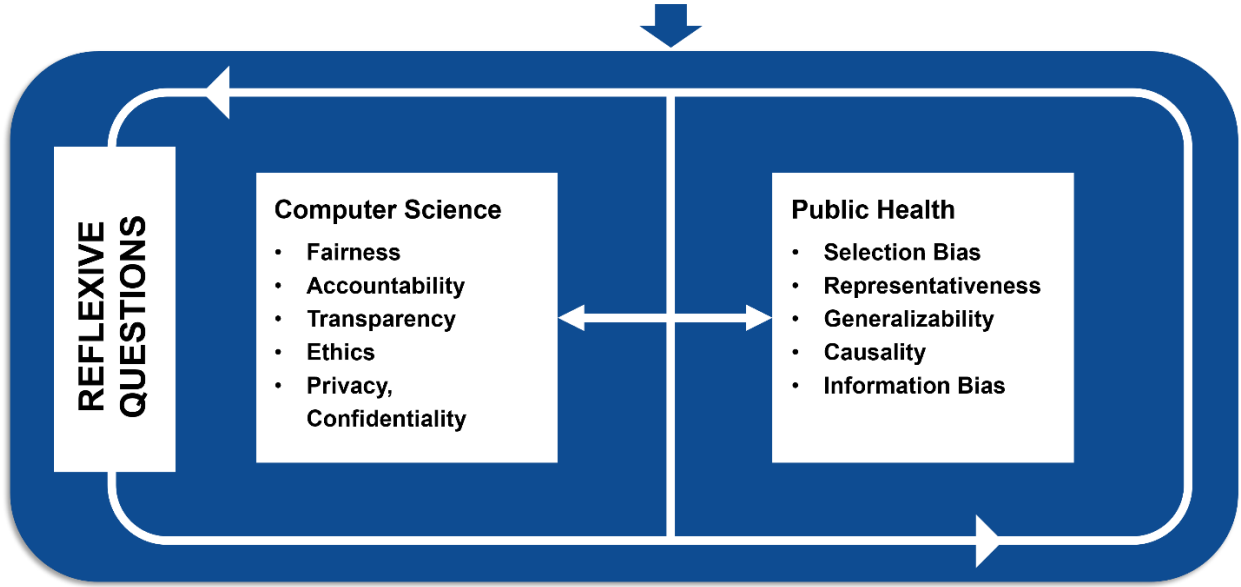

**Figure 1** Interdisciplinary conceptual framework for public health data equity. The framework integrates core principles from computer science and public health to create an implementable pathway to evaluate and address data equity. Reflexive questions ingrained in domain science promote feedback and interdisciplinary contextual grounding.

## 2. Data Equity Challenges across Public Health Research

Now, we demonstrate the existing challenges in performing *representative* scientific research through three illustrative examples from public health and medicine: (i) electronic health records (EHR) data, (ii) genome wide association study (GWAS), and (iii) pulse-oximeter medical device testing.

### 2.1. Data Inventory of Research Publications on Electronic Health Records

The promotion and growing adoption of EHR has driven significant advancements in healthcare and related research (Evans, 2016, Queen Elizabeth et al., 2023). However, stark geographical differences prevail in how EHR is used for health research. To objectively quantify this variation, we conducted a systematic search in Embase (Elsevier, 2025). Embase is a biomedical and pharmacological bibliographic database provided through Ovid (Wolters Kluwer, 2025), a widely used research platform that grants access to multiple biomedical databases. We performed a parallel supplementary analysis in an alternative database MEDLINE (National Library of Medicine, 2025) to further support our findings, and present results in the Supplementary Section S1.

We assess the volume and distribution of EHR-related research across geographic regions in the world by keyword-based search using the multi-purpose (.mp.) operator, which scans multiple fields in article records, including titles, abstracts, and indexing terms. This approach ensures that studies discussing EHR are captured even if they have not been formally categorized under a specific topic label within the database's indexing system. The .mp. search was performed using the query "(EHR or EMR or electronic health record* or electronic medical record* or electronic patient record*).mp.", ensuring broad coverage of relevant variations in nomenclature.

Although Embase covers publications from 1974 onward, this study focuses on EHR-related publications from 2014 to 2024. All records were retrieved on February 5, 2025. The .mp. keyword-based search identified 37,729 geographically tagged publications. To analyze the geographic distribution of these studies, we categorized them into 11 regions of interest by applying Emtree's geographic indexing terms, which ensured each study was classified based on its regional focus. Broader regions, such as Europe excluding the UK and North America excluding the US, were constructed by combining multiple geographic terms to ensure precise segmentation. For studies spanning multiple geographic regions, we assigned them to a "Multiple" category, while publications associated with geographic areas not covered by predefined categories were classified as "Other." To simplify visualization, these two groups were combined into a "Multiple & Other" category. The full search query details for both approaches, including Emtree terms used for each region, are provided in Supplementary Section S1.

Figure 2 presents trends in EHR-related research identified by the .mp. keyword search in Embase from 2014 to 2024. Figures 2(a) shows the annual count of EHR publications, and Figures 2(b) displays the proportional representation of EHR publications by 11 geographic regions in the world, along with a combined "Multiple & Other" category, with each bar representing a year and different colors within a bar indicating relative contributions from the different regions. The rightmost stacked bars provide a reference benchmark by displaying the global population distribution in these 11 regions, as reported in the 2024 Revision of World Population Prospects from the United Nations Department of Economic and Social Affairs (United Nations Department of Economic and Social Affairs Population Division, 2024).

As shown in Figures 2(a), the annual number of publications increased notably since 2017. In 2014, the .mp. search identified 2,469 publications, which increased to 4,564 in 2024, representing an 85% increase over the period. However, Figure 2(b) highlights that the distribution of research output across geographic regions remains highly skewed. The proportion of EHR studies attributed to the United States changed from 48.5% in 2014 to 39.7% in 2024, while contributions from the United Kingdom fluctuated between 6.1% and 10.2% over the years. Despite these shifts, research output from the US and the UK still accounted for roughly 50-60% of all EHR studies in Embase in 2024. This concentration of research in just two high-income countries remains disproportionate to their combined population share of only 5.1% of the global total. Meanwhile, populated regions such as Africa (18.6% of the global population), South Asia (24.2%), and China (17.5%), have remained underrepresented in EHR research. In 2024, studies from Africa constituted just 2.5% of Embase publications, while South Asia contributed 1.8%, and China accounted for 5.1%. A detailed numerical breakdown of EHR research representation by geographic region over time is provided in Tables S1.2 and S1.4, available in Supplementary Section S1.1.

**Summary:** Our analysis reveals a persistent discordance between the distribution of the global population and the geographic distribution of EHR research. While publications overwhelmingly concentrated in high-income countries, low- and middle-income regions remain severely underrepresented. South Asia, Africa, and China — comprising a significant share of the global population (nearly 60%) — continue to contribute only a small fraction of EHR-related publications (roughly 8-9%). This difference results from a multitude of reasons: late and limited adoption of EHR systems, weaker research infrastructure, lack of funding allocation for digitization of health data, and limited data accessibility, all of which restricts the global relevance and impact of EHR research advancements. Without targeted efforts to address these gaps, EHR-based research risks reinforcing existing challenges rather than mitigating them.

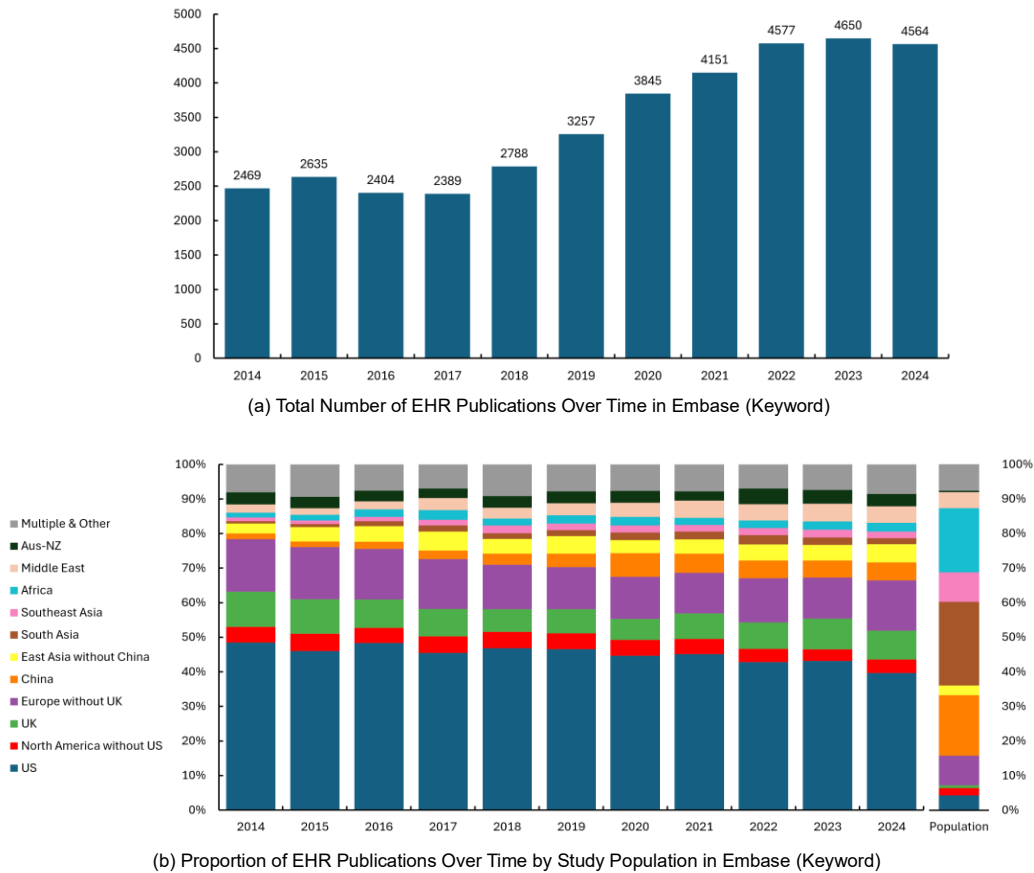

**Figure 2** Trends in EHR-related research indexed in Embase from 2014 to 2024, using the multi-purpose (.mp.) keyword search. (a) Annual number of EHR-related publications. (b) Proportional representation of EHR publications by study population region. In panel (b), the rightmost stacked bar represents the global population distribution estimated from the 2024 Revision of World Population Prospects, published by the Population Division of the United Nations Department of Economic and Social Affairs (United Nations Department of Economic and Social Affairs Population Division, 2024).

## 2.2. Genome-Wide Association Studies

Genome-Wide Association Studies (GWAS) have significantly advanced our understanding of genetic contributions to various diseases and traits (Uffelmann et al., 2021). However, concerns around unequal availability of genomic data among different ancestral populations have been repeatedly raised (Fatumo et al., 2022, Fitipaldi and Franks, 2022, Troubat et al., 2024, Corpas et al., 2025). Fatumo et al. (2022) reported that, as of June 2021, individuals of European descent accounted for 86% of all genomics studies, an increase from 81% that was noted in 2016, while the representation of non-European populations has remained disproportionately low. This persistent imbalance limits the generalizability of GWAS findings, as results derived primarily from European populations may not optimally apply to other populations despite using the most sophisticated cross-ancestry transfer learning methods (Wang et al., 2020, Chen et al., 2025, Zhu et al., 2025), reducing their clinical and biomedical relevance in global contexts.

We analyzed the ancestry dataset from the NHGRI-EBI GWAS Catalog (Cerezo et al., 2025) as of May 13, 2025, to examine the representation of ancestry populations in GWAS research. This dataset includes records spanning more than two decades, from March 10, 2005, to April 25, 2025, providing a comprehensive overview of the evolution of GWAS representation over time. Each record in the dataset represents a distinct ancestry population reported within a GWAS study, allowing for detailed analysis of population representation across studies. The original dataset contains 198,017 records, where each record represents a combination of study metadata, including PubMed ID, first author's name, publication date, stage of the GWAS study (initial or replication, where initial refers to the discovery phase and replication validates findings in independent samples), ancestry description for different study stages, number of individuals in the sample, ancestry category, country of origin of the individuals, country of recruitment, and additional ancestry-related descriptors relevant to the sample. A detailed account of the data acquisition, preprocessing, and visualization methods used in this example is provided in Supplementary Section S2.

To ensure uniformity in ancestry classification, we harmonized the ancestry categories to address variations in labeling across studies. Study populations were classified into ten broad ancestry groups: European, East Asian, South Asian/Other Asian, African, Hispanic/Latino, Greater Middle Eastern, Oceanic, Other, Multiple, and Not Reported. Studies involving multiple ancestry groups were assigned to the "Multiple" category, while entries labeled as "Not Reported" were excluded from further analysis.

We also implemented a filtering process to identify and consolidate redundant entries that could inflate participant counts. Duplicates were defined as records sharing the same PubMed ID, study stage, ancestry category, and participant count, and these entries were merged accordingly. In addition to standard deduplication, many studies contained overlapping samples across records with the same PubMed ID, leading to repeated counting of the same individuals. The nature of these overlaps varies across studies, requiring tailored curating approaches to accurately estimate their unique contributions to the number of genomes.

The cleaned dataset was used to generate visual representations of GWAS ancestry disparities, presented in Figure 3. The upper left section of Figure 3 illustrates the cumulative number of individuals studied over time across nine different ancestry groups, totaling approximately 1.83 billion individuals across study samples between March 10, 2005, and April 25, 2025. The lower left section depicts the proportional representation of each ancestry group relative to the total GWAS sample size at each time point. To contextualize these findings, the right vertical bar of Figure 3 presents the 2025 global population distribution based on data from Worldometer (Worldometer, 2025), which reports region-level estimates that more closely align with the ancestry categories used in GWAS studies. Mapping details for aligning population regions with ancestry categories are provided in Supplementary Section S2.4. For comparative purposes, we assume that the regional population proportions have remained relatively stable from 2005 to 2025.

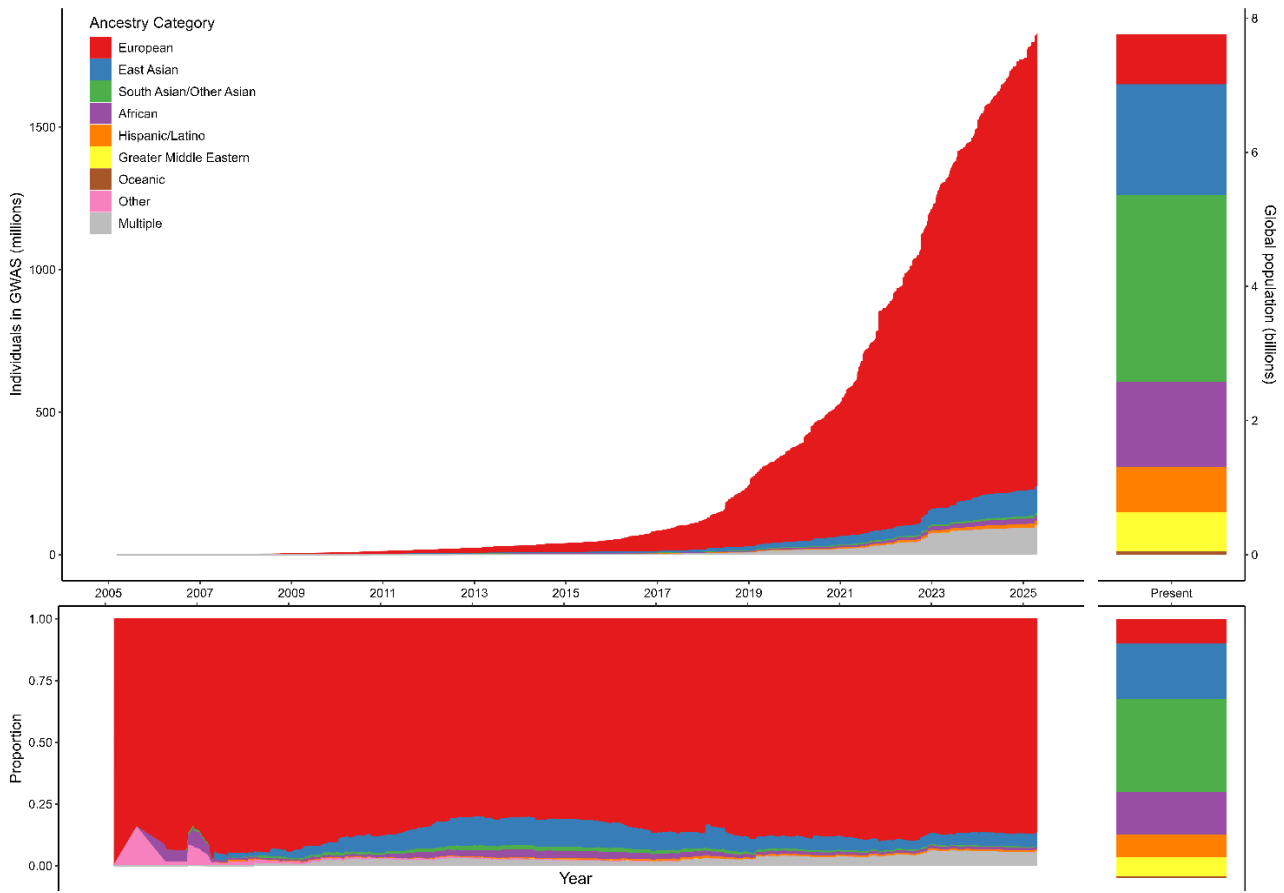

**Figure 3** The upper-left panel shows the cumulative number of individuals included in GWAS studies over time by ancestry group, and the lower-left panel displays their proportional representation. Both panels are based on data from the NHGRI-EBI GWAS Catalog (Cerezo et al., 2025). Between March 10, 2005, and April 25, 2025, these studies collectively reported approximately 1.83 billion individuals (n = 1,825,314,807) across study samples. The rightmost bar presents the global population distribution by ancestry group in 2025, based on region-level estimates from Worldometer (2025).

This quantification highlights persistent disparities in genomic research. As of April 25, 2025, among the approximately 1.83 billion individuals represented in the GWAS dataset, 87.0% were of European ancestry, despite representing only 9.0% of the estimated global population. In contrast, individuals of African ancestry, who constitute 15.5% of the estimated global population, were represented in just 1.1% of GWAS studies. Similarly, the South Asian/Other Asian ancestry group, which has the largest estimated population share at 33.8%, contributed only 0.4% to GWAS datasets. The East Asian ancestry group, representing 20.0% of the estimated global population, accounted for 5.1% of GWAS participants, also indicating a notable underrepresentation relative to its population size. Other ancestries, including Greater Middle Eastern, Oceanic, and Hispanic/Latino populations, remained marginally represented, with proportions below 1% despite their substantial global presence. A detailed breakdown of GWAS representation as of April 25, 2025, and estimated population proportions for ancestry groups in 2025 provided in Supplementary Table S2.1.

**Summary:** The severe underrepresentation of non-European ancestries in GWAS highlights the pressing need for more inclusive research practices in genetics. This data gap impedes the identification of ancestry-specific genetic variants that could inform therapeutic targets and precision medicine. Addressing these gaps requires intentional efforts to expand representation in GWAS datasets, ensuring that genomic research produces findings that are both scientifically robust and equitably distributed (Peterson et al., 2019, Dolan et al., 2024).

## 2.3. Racial Differences in Pulse-Oximetry Measures

Pulse oximetry is a widely used noninvasive technology in healthcare that estimates arterial oxygen saturation ($SpO_2$) as a measure of true arterial oxygen saturation ($SaO_2$) (Feiner et al., 2007, Fawzy et al., 2022). It operates by analyzing the differential absorption of red and near-infrared light by oxyhemoglobin and deoxyhemoglobin in capillary blood (Chesley et al., 2022, Holder and Wong, 2022). This technology has become indispensable in critical care settings, including emergency departments, operating rooms, and intensive care units (ICU), where continuous monitoring of oxygen levels is vital for patient care (Bothma et al., 1996, Adler et al., 1998, Fawzy et al., 2022, Valbuena et al., 2022b). During the COVID-19 pandemic, pulse oximeters played a crucial role in determining patient care, as $SpO_2$ thresholds were frequently used for triaging and guiding treatment decisions (Holder and Wong, 2022).

However, the accuracy of pulse oximetry has faced significant scrutiny for poor performance in users with darker skin tones. A critical example of this issue is occult hypoxemia, a condition in which pulse oximeters fail to detect dangerously low oxygen levels, leading to overestimated oxygen saturation readings (Ward and Katz, 2022). Studies revealed that this phenomenon disproportionately affects individuals with dark skin pigmentation, a long-standing issue first identified in the literature more than 30 years ago (Jubran and Tobin, 1990, Zeballos and Weisman, 1991, Bickler et al., 2005, Henry et al., 2022, Valbuena et al., 2022a, Valbuena et al., 2022c, Barreto and Moynihan, 2023, Dempsey et al., 2024, Reep et al., 2024). Feiner et al. (2007) reported that the oximetry bias is most pronounced in individuals with dark skin, moderate in those with intermediate skin tones, and minimal in those with light skin. Sjoding et al. (2020) found that Black patients experienced nearly three times the rate of occult hypoxemia compared to White patients. Similar findings have been reported for other racial and ethnic groups, including Asian and Hispanic patients, highlighting the pervasive impact of skin pigmentation on pulse oximeter performance (Wong et al., 2021, Fawzy et al., 2022, Gottlieb et al., 2022).

These inaccuracies have had severe clinical consequences, particularly during the COVID-19 pandemic, when pulse oximetry played a critical role in triaging and managing respiratory failure. Overestimation of oxygen saturation by pulse oximeters delayed recognition of treatment needs and hospital admissions for Black and Hispanic patients (Fawzy et al., 2022, Sudat et al., 2023). Holder and Wong (2022) found that clinicians' reliance on pulse oximetry to triage and manage COVID-19 treatment disproportionately increased ICU hospitalization and mortality risks among non-White patients. Similarly, Savorgnan et al. (2023) noted that in pediatric COVID-19 cases, Black patients face a higher risk of occult hypoxemia, further highlighting the inequities in critical care.

Beyond oxygenation errors, racial disparities in $SpO_2$ measurements extend to tertiary outcomes such as cardiovascular assessment. Ketcham et al. (2024) found that $SpO_2$ measurements underestimated Fick cardiac output compared to $SaO_2$ in Black patients, potentially leading to misguided treatment decisions. This inaccuracy could affect critical interventions such as inotropes and mechanical circulatory support, further compounding the risks of inequitable healthcare outcomes.

**Summary:** Systematic measurement biases in pulse oximetry not only affect individual diagnoses and treatment, but also introduce structural inaccuracies into population-level health datasets. These inaccuracies can distort epidemiological findings, misinform public health interventions, and reinforce disparities in critical care access and outcomes. Numerous studies emphasize the need for pulse oximeter validation across a broader range of patient populations and advocate for manufacturers to develop objective methods to account for skin pigmentation when estimating arterial oxygen saturation (Wong et al., 2021, Henry et al., 2022, Holder and Wong, 2022, Jamali et al., 2022, Moore et al., 2022, Okunlola et al., 2022, Palmer, 2022, Barreto and Moynihan, 2023, Reep et al., 2024). In response to these growing concerns, the U.S. Food and Drug Administration (FDA) has taken steps to evaluate and improve pulse oximeter accuracy in diverse populations. On November 16, 2023, the FDA published a discussion paper outlining strategies to enhance the quality of premarket studies and evaluation methods for pulse oximeters (U.S. Food and Drug Administration, 2023a). This paper emphasizes the importance of considering patient skin pigmentation as well as self-reported race and ethnicity. Furthermore, to address these critical issues and incorporate expert input, the FDA convened two meetings of the Center for Devices and Radiological Health Anesthesiology and Respiratory Therapy Devices Panel (U.S. Food and Drug Administration, 2023b). These regulatory efforts reflect a growing recognition of the need for more inclusive validation protocols, inclusion of diverse groups in medical device clinical trials, reporting performance across sub-groups.

These three examples demonstrate that significant inequity and bias permeate various data domains in public health, profoundly influencing health outcomes. Now we answer the question, what tools and frameworks currently exist?

## 3. Contemporary Literature

### 3.1. Data Equity in Computer Science

The discussion of data equity within the computer science literature is still in its very nascent phase, with the term itself having emerged explicitly during the global COVID-19 pandemic (Jagadish et al., 2021). While the concept itself is not new, its formal recognition and structured examination within the field has gained momentum only recently.

Jagadish et al. (2021) provided a foundational framework for understanding data equity by categorizing its challenges into four key dimensions:

(1) **Representation equity:** ensuring visibility and inclusion of historically underrepresented groups in datasets.

(2) **Feature equity:** enhancing the ability to link datasets to effectively discover and quantify inequities.

(3) **Access equity:** promoting equitable and participatory access to data and data products, regardless of expertise or domain.

(4) **Outcome equity:** monitoring and mitigating unintended consequences of deployed systems on affected groups, both directly and indirectly.

Building on this framework, Jagadish et al. (2022) expanded this discussion by advocating for a holistic, socio-technical approach to data equity, emphasizing that data equity must be integrated from the inception of artificial intelligence (AI) and machine learning (ML) system design. Aligning with this perspective, Leslie et al. (2022) emphasized that equity considerations should guide decisions from the earliest stages of project planning, even before data collection begins, ensuring that principles of fairness underpin the entire data life cycle. Similarly, Schwartz et al. (2022) argued that equity should remain a continuous focus throughout the AI development life cycle, advocating for an iterative process that revisits equity concerns as systems evolve.

Institutional efforts have also shaped the discourse. World Economic Forum (2023) formalized the integration of equity through three key stages across the data life cycle: input data equity (ensuring representation and feature equity at the data collection stage), algorithmic data equity (addressing representation, feature, and access equity during model development), and output data equity (focusing on access and outcome equity in deployed AI systems). The World Economic Forum's 2024 report further operationalized this framework by identifying ten core characteristics of data equity — such as accountability, accessibility, trust, and value — that provide a foundation for evaluating and implementing equitable practices across diverse data systems (World Economic Forum, 2024).

Recent contributions have expanded this foundation by addressing data equity through broader lenses of governance and public engagement. Gennari (2024) linked data equity to global data governance and official statistics, underscoring the need for equitable access, inclusive representation, and fair distribution of benefits in an era dominated by private-sector data proliferation. Complementing this institutional perspective, Kennedy et al. (2024b) examined how the public understands and articulates fairness and equity in data usage within UK public services, highlighting the importance of embedding social values into governance frameworks to support equitable outcomes.

Collectively, these studies underscore the critical need to integrate data equity principles at every stage of technological development. By embedding equity from data collection to model deployment, AI systems can become not only more effective but also fair, inclusive, and reflective of diverse perspectives.

## 3.2. Data Equity in Public Health

In public health, the concept of data equity has been applied primarily in specific contexts rather than as a cohesive framework. Early mentions of equity related to data include Guidotti (2002), who highlighted the lack of detailed data for assessing asbestos-related health risks among workers. More recently, Ponce et al. (2015) emphasized the need for representative population health data, particularly from minority groups, as an essential step toward data equity. Over time, the term has been associated with diverse topics in public health, including food insecurity (Cao et al., 2024), climate change (Kennedy et al., 2024a), precision medicine (Caton and Haas, 2024), and reproductive health (Heyrana et al., 2023). A consistent theme across these areas is the underrepresentation of marginalized populations in data collection, which limits the applicability and fairness of public health interventions.

Efforts to integrate data equity into research and data science practices have focused on improving data collection practices (O'Neil et al., 2021, Wilson et al., 2024), training staff to recognize biases (Duerme et al., 2021), and developing frameworks to guide equitable research methodologies (Mays et al., 2022, Sapat et al., 2022). For example, López-Cevallos et al. (2023) advocated prioritizing communities of color in national survey cycles to enhance the utility of data for public health actions. However, privacy concerns have emerged in contexts such as use of wearable devices and mobile technologies for health surveillance, as excessive monitoring risks harming vulnerable populations by enabling behavioral misinterpretation, extending risk through data immortality, and allowing reidentification and misuse of personal health information (Kilgallon et al., 2022). More broadly, concerns about data misuse extend beyond digital health technologies, underscoring the importance of ensuring transparency and securing informed consent in all forms of data collection. These measures are particularly critical for building trust among participants, especially within marginalized communities (Mays et al., 2022, Sapat et al., 2022).

The link between data equity and health equity became increasingly evident during the COVID-19 pandemic. Marginalized populations faced heightened vulnerabilities due to inequities in healthcare systems and data collection practices (Sapat et al., 2022, Kauh et al., 2023). Meanwhile, there are significant deficiencies in data reporting and availability, particularly in low- and middle-income countries (Durieux and Naik, 2022). Post-pandemic, calls for equitable data practices have grown louder, with recommendations ranging from disaggregated data collection to targeted investments in digital infrastructure and workforce training (Ponce et al., 2015, Lee et al., 2023, Lam‑Hine et al., 2024).

Despite the growing discourse on data equity in both computer science and public health, the literature remains scattered and lacks a **unified, action-oriented framework.** Apart from the World Economic Forum's white paper on data equity (World Economic Forum, 2024), there is no comprehensive roadmap for implementing data equity in practice. This creates a critical disconnect, leaving the vital, bidirectional relationship between data equity principles and public health methodologies largely unexplored and undefined.

Next, we attempt to bridge this gap by integrating key principles from both computer science and public health to develop a structured approach for operationalizing data equity within the public health data science workflow.

# 4. Overview of Core Concepts and Metrics Related to Data Equity

The disciplines of computer science and public health approach data analysis from distinct perspectives. Computer science often emphasizes "how" to analyze given data by building scalable algorithms for prediction, optimization, and deployment in real-world systems. Statistics and public health often provide explanations of "why" insights drawn from a limited sample can be validly extended to broader populations, emphasizing generalizability, causal reasoning, and population-level inference (Lau et al., 2022). These differences shape how each field conceptualizes and operationalizes data equity. At the intersection of these disciplines, the emerging area of public health data science integrates computational techniques with causally defensible/reasoned inferential frameworks to address data equity challenges across the data life cycle. The following sections examine core concepts that form pillars for data equity, beginning with foundational ideas in Computer Science and followed by concepts from Public Health.

## 4.1. Data Equity-Related Concepts Rooted in Computer Science

Literature in computer science provides the key principles and technical frameworks that drive AI and ML innovations. These frameworks have contributed to many societal processes, offering efficiency and innovation across domains such as education, healthcare, and manufacturing industry (Shailaja et al., 2018, Korkmaz and Correia, 2019, Whittlestone et al., 2021). However, these innovations also pose challenges related to data equity, as they typically rely on incomplete or biased samples of the target population. This can unintentionally reinforce systemic disparities (Pessach and Shmueli, 2022). Moreover, AI and ML methods often treat data as given and for most of the frameworks there is a very limited focus on statistical inference and generalizability (Schwartz et al., 2022). To address some of these issues, principles of fairness, accountability, transparency, and ethics (FATE) have been integrated into the design and application of AI/ML systems.

FATE serves as a *conceptual* framework to evaluate and improve AI's alignment with societal values, as synthesized by Boyd (2021) and Memarian and Doleck (2023). *Fairness* involves ensuring impartial treatment and mitigating biases in both data and algorithms. *Accountability* requires that systems be auditable and that their outcomes as well as inner workings can be traced and explained. *Transparency* recommends models to be interpretable, understandable and communicable to relevant parties. Finally, *ethics* encompasses the broader principles of responsibility and trustworthiness, guiding the development and deployment of AI systems in ways that respect human rights and societal norms. Together, these principles provide a foundation for creating AI systems that are both accurate and socially responsible.

Beyond FATE, *privacy* and *confidentiality* play a crucial role in fostering trust and protecting individual rights in AI-driven systems (Harishbhai Tilala et al., 2024). Protecting privacy extends beyond mere compliance — it is about creating a balance between data utility and the safeguarding of sensitive information. Similarly, maintaining confidentiality ensures that personal data entrusted to researchers or AI systems is handled responsibly and ethically. As AI applications scale and data-driven decision-making become more pervasive, incorporating privacy and confidentiality into AI frameworks is essential to upholding the values of inclusivity, security, and respect for all individuals.

These principles are central to addressing data equity challenges in AI-driven decision-making. Now, we delineate each of these key concepts and discuss how these computational frameworks shape equitable data practices and inform AI governance strategies.

**Fairness**

***Existing Literature.*** Fairness in ML refers to ensuring that ML models make decisions that are impartial and equitable across subgroups, though its interpretation varies depending on the societal or application context (Mulligan et al., 2019). The concept of fairness has been examined through both conceptual and mathematical frameworks. Conceptually, it is often divided into two broad categories: individual fairness and group fairness (Mulligan et al., 2019). Individual fairness, as defined by Dwork et al. (2012), ensures that similar individuals receive similar treatment, reflecting equity at a granular level. In contrast, group fairness focuses on achieving comparable outcomes or treatment across predefined demographic groups, ensuring broader parity in algorithmic decisions (Pedreschi et al., 2008, 2009).

Mathematically, fairness has been formalized through various frameworks (Dwork et al., 2012, Feldman et al., 2015, Hardt et al., 2016, Kleinberg et al., 2016, Zafar et al., 2017), where two widely used definitions — demographic parity and equalized odds — are particularly influential and were formalized clearly by Agarwal et al. (2018). We consider a binary classification setting where each observation is a triple $(X, A, Y)$, with $X \in \mathcal{X}$ representing a feature vector, $A \in \mathcal{A}$ a protected attribute (such as race or gender), and $Y \in \{0, 1\}$ the binary outcome. The goal is to learn a classifier $h: \mathcal{X} \to \{0, 1\}$ that predicts $Y$ from $X$, ideally without unfair influence from $A$. Demographic parity requires the prediction $h(X)$ to be statistically independent of the protected attribute $A$. That is, a classifier satisfies demographic parity if

$$P(\,h(X) = \hat{y} \mid A = a\,) = P(h(X) = \hat{y}) \text{ for all } a, \hat{y},$$

which is equivalent to

$$E[\,h(X) \mid A = a\,] = E[h(X)] \text{ for all } a,$$

when $\hat{y} \in \{0, 1\}$. This ensures equal rates of positive predictions across subgroups, but it may result in poor classification performance when outcome rates differ by group. In particular, demographic parity may allow random predictions for underrepresented groups, and it precludes perfect classification when the outcome is correlated with the protected attribute.

Equalized odds refines this by conditioning on the true outcome $Y$, requiring the prediction to be independent of $A$ given $Y$. Formally, a classifier satisfies equalized odds if

$$P(\,h(X) = \hat{y} \mid A = a, Y = y\,) = P(h(X) = \hat{y} \mid Y = y) \text{ for all } a, y, \hat{y},$$

which is equivalent to

$$E[\,h(X) \mid A = a, Y = y\,] = E[h(X) \mid Y = y] \text{ for all } a, y,$$

when $\hat{y} \in \{0, 1\}$.

However, these metrics can be often mutually incompatible, requiring researchers to weigh societal goals and ethical considerations when selecting the appropriate measure (Srivastava et al., 2019). As Friedler et al. (2021) argue, each fairness definition reflects a distinct worldview, and designing fair algorithms necessitates explicitly acknowledging how observed data relate to the idealized constructs underlying decision-making.

Another notable formalization is counterfactual fairness, which defines a predictor as fair if its output for an individual would remain the same in a hypothetical scenario where the individual's sensitive attribute, such as race or gender, was different, while all other relevant characteristics are held constant (Kusner et al., 2017). Formally, a predictor $\hat{Y}$ is counterfactually fair if under any $X = x$ and $A = a$,

$$P(\hat{Y}(a) = y \mid X = x, A = a) = P(\hat{Y}(a') = y \mid X = x, A = a)$$

for all $y$ and for any value $a'$ attainable by $A$. Here $\hat{Y}(a)$ denotes the counterfactual prediction had the protected attribute $A$ taken the value $a$, under the assumption that all other features $X$ remain as observed and are unaffected by this change. This formulation captures the idea that changing $A$ while holding features not causally dependent on $A$ constant should not affect the predicted outcome. Expanding on this causal perspective, Nabi and Shpitser (2018) proposed a causal framework where fairness is formalized as the absence of certain path-specific effects (PSEs) from sensitive attributes to outcomes. To enforce fairness, they constructed a new distribution that is closest to the observed data in terms of Kullback–Leibler (KL) divergence, i.e., the distribution that minimally deviates from the empirical data while satisfying fairness constraints on the magnitude of unfair PSEs. Their approach captures intuitive notions of fairness, for instance ensuring that gender does not directly influence hiring decisions, while still allowing effects mediated through legitimate applicant characteristics. Alongside these theoretical developments, fairness-aware algorithms have emerged across different stages of the ML pipeline: pre-processing methods modify data to reduce bias (e.g., reweighting or balancing); in-processing techniques incorporate fairness constraints into the learning process; and post-processing adjusts predictions or thresholds to improve group parity (Raza, 2022).

Despite these efforts, guaranteeing fairness requires understanding societal dynamics and navigating trade-offs between conflicting definitions, model accuracy, and complexity, as highlighted by Holstein et al. (2019). Survey studies, such as those by Pessach and Shmueli (2022) and Caton and Haas (2024), have provided comprehensive reviews of fairness metrics and mitigation strategies, offering practical guidance for integrating fairness into real-world applications.

***Fairness as a Pillar for Data Equity.*** In the context of data equity, fairness focuses on mitigating algorithmic biases in data systems that stem from biased training data, incorrect assumptions or proxies, and gaps in algorithm design. The theory of fairness encourages that design of data systems actively promote collection and use of data from populations that are often missing, such as low income, minority, differently abled, or geographically isolated groups (Rajkomar et al., 2018, Raza et al., 2023).

Despite growing awareness of these imperatives, the adoption of fairness metrics in practice remains limited. A recent review by Rountree et al. (2025a) found that while performance comparisons across sensitive features, such as sex or race, were occasionally reported, explicit use of fairness metrics in clinical risk prediction models was largely absent among high-impact studies on cardiovascular disease and COVID-19. To bridge the gap between theory and practice, the authors offered a four-part roadmap for changing practice: Interpret, Implement, Connect, and Collect (I2C2), which emphasizes the need to justify the use of sensitive features, update reporting guidelines to include fairness metrics, connect algorithmic fairness research to clinical practice, and prioritize inclusive data collection. They also summarized widely used fairness metrics such as demographic parity and equalized odds, reinforcing the need for standardized approaches to evaluating equity in precision health.

Incorporating fairness into data equity frameworks also aligns with broader ethical principles, enabling the development of data systems that are responsive to the unique challenges faced by underrepresented groups. Embedding fairness into every stage of the data life cycle — such as ensuring inclusive data collection, equitable model design, and careful interpretation — helps ensure that algorithms avoid unintentional errors due to missing or unrepresentative data. After data analysis is complete, attention to fairness promotes inclusive access to and use of results across populations.

**Accountability**

***Existing Literature.*** Memarian and Doleck (2023) defined accountability as a set of preventative or corrective measures that hold the creators, designers, and users of AI systems responsible for the outcomes their algorithms produce. Ensuring accountability in AI systems requires a clear understanding of who holds responsibility at different stages of the data life cycle.

One fundamental challenge in ensuring accountability is the *many hands* problem, where responsibility becomes fragmented among multiple parties involved in designing, operating, and deploying AI systems (Nissenbaum, 1996). This diffusion of responsibility can obscure clarity about who should be held accountable for adverse outcomes, as noted by De Laat (2018). The complexity of AI ecosystems further complicates accountability, making it crucial to establish explicit mechanisms that delineate responsibility across involved entities.

Recognizing the distributed nature of AI governance, World Economic Forum (2023) identified three key groups in AI accountability: (1) those responsible for driving and governing AI, including AI-creating and AI-using organizations as well as policymakers; (2) those impacted by or who are the end users of AI systems, particularly the public and communities; and (3) those who bridge concerns between these two groups, such as civil society organizations that focus on capacity-building and advocating for inclusive AI governance. Addressing accountability requires engaging all three groups to ensure that AI governance mechanisms reflect not only technical and regulatory considerations but also societal needs and ethical concerns.

Another major debate in AI accountability centers on whether humans alone should bear responsibility or if AI systems themselves should share it. While some scholars advocate for human-centered accountability (Kong et al., 2023), others propose a shared model that considers AI systems as active participants in decision-making processes (Li et al., 2022), for example by embedding fairness constraints directly into algorithmic design. These discussions highlight the need for robust frameworks that explicitly define accountability at every stage of the data life cycle, ensuring transparency and fairness in AI-driven decision-making.

***Accountability as a pillar for data equity.*** Accountability requires those with influence over data systems—such as researchers, funders, policymakers, and technology developers—to take responsibility for the design, implementation, and consequences of data collection and analysis. While many groups are affected by data practices, the responsibility to identify and mitigate harms falls especially on those with decision-making power. This includes engaging relevant parties during the design process, documenting key decisions, and establishing mechanisms to address unfair or harmful outcomes. Embedding accountability throughout the data life cycle ensures that responsibility for data decisions is visible, traceable, and tied to those with the power to act. (World Health Organization Regional Office for Europe, 2019, Bartlett et al., 2024).

Incorporating accountability into data equity frameworks also strengthens public trust in data systems by ensuring that decision-makers remain answerable for outcomes. Establishing clear lines of responsibility for identifying, investigating, and remedying harmful impacts creates a culture where corrective action is expected and enforced.

**Transparency**

***Existing Literature.*** Transparency refers to the clarity and openness with which an algorithm's functioning, decision-making processes, and outcomes are made understandable and accessible to relevant partners, such as patients and healthcare providers (Memarian and Doleck, 2023). Diakopoulos (2016) identified five key aspects of transparency: human involvement, data, model, inference, and algorithmic presence. Nyrup (2022) further distinguished three types of transparency in ML based on value considerations: epistemic transparency, which concerns the uncertainties and justifications behind model decisions; retrospective value transparency, which examines the ethical and value-based influences on past decisions; and prospective value transparency, which evaluates the potential consequences and justifications for future decisions.

Achieving transparency, however, is not without challenges. De Laat (2018) highlighted "opacity" as a significant barrier, particularly in complex systems where the underlying processes may be difficult to interpret or disclose fully. To address this, total transparency was advocated, emphasizing that all phases of algorithmic processes should be accessible for public scrutiny, as a means to restore accountability and build trust. Zhou and Kantarcioglu (2020) further proposed two critical aspects for transparency-aware ML: "know-how," which focuses on designing human-intelligible decision objects, and "know-that," which seeks to understand the deeper justifications behind decisions.

Recent advancements in explainable AI have introduced methodologies to enhance transparency, particularly in complex models like deep learning, where decision-making processes are often opaque. Xu et al. (2019) outlined two primary approaches: transparency by design, which involves constructing inherently interpretable models by making model structures, parameters, and training algorithms more understandable, and post-hoc explainability, which generates explanations for decisions after they are made through analytic statements, visualizations, or example-based reasoning.

One formal approach to post-hoc transparency is Quantitative Input Influence (QII), which measures how much a model's output depends on a specific input feature (Datta et al., 2016). For a model $\mathcal{A}$, the QII of feature $i$ is defined as

$$\iota^{\mathcal{A}}(i) = Q_{\mathcal{A}}(X) - Q_{\mathcal{A}}(X_{-i}U_i),$$

where $Q_{\mathcal{A}}(X)$ is a quantity of interest, such as the probability of a positive prediction, and $X_{-i}U_i$ denotes the input vector where feature $i$ has been randomly pertubed. This quantifies the marginal effect of input $i$ on the model's behavior, offering a concrete way to interpret which inputs influence outputs and to what extent, thereby operationalizing transparency in complex models.

***Transparency as a pillar for Data Equity.*** Transparency is a fundamental pillar of data equity, ensuring that decision-making processes are open, understandable, and subject to scrutiny. It allows relevant partners, such as patients and healthcare providers, to assess not only how decisions are made but also whose interests they serve and whether they align with ethical and legal standards. Without transparency, it becomes difficult to evaluate whether an algorithmic system applies appropriate and unbiased criteria, making trust in its outputs uncertain (Shook et al., 2018). Transparent data practices facilitate well-documented and interpretable decision-making, enabling researchers, policymakers, and the public to assess the rationale behind algorithmic decisions, detect potential biases, and implement corrective measures that promote fairness and inclusivity.

**Ethics**

***Existing Literature.*** Ethics encompasses principles of right and acceptable conduct, shaped by societal norms and values, that guide decision-making and behavior, particularly in situations where values conflict or lack clarity (Stahl et al., 2016). In the context of modern data practices, ethics addresses pressing concerns such as data privacy, security, and the potential for misuse, especially as more personal information is collected and processed online. Ethical considerations have become increasingly central to discussions about the development and deployment of AI and ML technologies, particularly in critical areas like healthcare (Harishbhai Tilala et al., 2024).

The rapid advancement of AI has spurred the creation of ethical guidelines aimed at ensuring responsible technological development and application (Hagendorff, 2020). These guidelines often outline principles that technology developers are encouraged to follow. However, the implementation of these guidelines varies significantly across institutions and societies, reflecting differences in ethical priorities and governance structures. Furthermore, the global conversation on AI ethics reveals notable imbalances, with underrepresentation from regions such as Africa, South and Central America, and Central Asia, which highlights inequities in the international discourse (Jobin et al., 2019).

Efforts to embed ethics into AI development frequently emphasize fairness, accountability, and inclusivity. Aler Tubella et al. (2022) underscored how technical decisions made during data pre-processing and post-processing are inherently value-laden and can influence fairness outcomes. Schwöbel and Remmers (2022) proposed a dynamic fairness model that integrates ethical principles with measurable fairness metrics and short- and long-term impact assessments, offering a structured approach to ethical algorithmic decision-making.

The concept of Responsible AI (RAI) has emerged as a critical governance framework to address the societal risks of AI, including unethical and irresponsible usage (Dignum, 2019, Bak, 2022). Anagnostou et al. (2022) outlined practical steps for implementing RAI, such as establishing ethical guidelines, enhancing accountability mechanisms, and fostering a shared ethical language in AI deployment. These strategies underscore the importance of deliberate, structured approaches to ensuring that technological innovations serve society responsibly and equitably.

***Ethics as a Pillar for Data Equity***. Ethical stewardship is essential to ensure that data science resources and innovations benefit all communities equitably. As a core pillar of data equity, ethics guides the responsible development and deployment of data systems to prevent harm, ensure inclusion, and align data practices with broader human and societal values. Even when ethical frameworks exist, data systems may still amplify inequities if principles are not meaningfully applied (Jobin et al., 2019). Hagendorff (2020) emphasized the importance of enforceable measures within ethical guidelines to ensure accountability and safeguard against misuse. Embedding ethics across the data life cycle provides a clear and easily recognizable language for making decisions about data practices.

**Privacy and Confidentiality**

*Existing Literature.* Privacy refers to the set of protections individuals have over their personal information, while confidentiality is the obligation of those who receive this information to safeguard it (Hulkower et al., 2020). These concepts are particularly significant in public health, especially in the context of electronic health records (EHRs), where ensuring the protection of sensitive data is crucial (Terry and Francis, 2007, Harman et al., 2012, Bani Issa et al., 2020). In the United States, the Health Insurance Portability and Accountability Act of 1996 (HIPAA) established federal standards to protect the privacy of personal health information. Under HIPAA, specific policies, procedures, and guidelines are defined to ensure that entities handling patients' identifiable health information uphold strict privacy and confidentiality standards (Office for Civil Rights, 2002).

However, implementing privacy and confidentiality measures often comes at a cost. Ness and Joint Policy Committee (2007) found that the HIPAA Privacy Rule introduced uncertainties, increased costs, and delays in human subjects research, highlighting the potential unintended consequences of stringent privacy regulations. Additionally, one of the central challenges in privacy-preserving methods is the trade-off between privacy protection and data utility. Techniques such as anonymization and differential privacy are designed to safeguard individual information, but they can reduce the granularity and accuracy of datasets, particularly for small demographic or geographic subgroups (Pujol and Machanavajjhala, 2021).

A widely used formalization of privacy is differential privacy (DP), which offers a mathematical guarantee that the inclusion or exclusion of any individual in a dataset does not substantially affect the analysis outcome. Specifically, a randomized algorithm $\mathcal{A}$ satisfies $\epsilon$-differential privacy if for any two datasets $D_1$ and $D_2$ differing in at most one entry, and for any subset of outputs $S$:

$$P[\mathcal{A}(D_1) \in S] \leq \exp(\epsilon) \times P[\mathcal{A}(D_2) \in S].$$

Here, $\epsilon$ represents the privacy budget, which quantifies the allowable privacy loss. Lower values of $\epsilon$ indicate stronger privacy. The exponential function provides a multiplicative bound on how much the presence or absence of a single individual can influence output probabilities, and its mathematical properties allow for precise control and composition across multiple queries. This formulation enables precise control over information leakage and has become a cornerstone in privacy-preserving data analyses, including applications in public health surveillance and census data release (Dwork, 2008).

The balance between privacy and data accuracy is particularly evident in geographical datasets. Waller (2024) illustrated how differentially private methods introduce noise to prevent re-identification, yet this can disproportionately impact smaller regions or minority populations. Global differential privacy applies fewer constraints at the subgroup level, maximizing overall privacy but limiting precision in localized analyses. In contrast, subgroup differential privacy improves accuracy for smaller groups but can reduce privacy protections. This trade-off is critical in applications such as disease mapping and resource allocation, where geographical precision is essential for equitable decision-making.

Further complicating these challenges, Rocher et al. (2019) demonstrated that even heavily anonymized datasets remain vulnerable to re-identification. Their study found that 99.98% of Americans could be correctly re-identified using only 15 demographic attributes, raising concerns about the effectiveness of traditional anonymization techniques. As data sharing expands, particularly in large-scale public health surveillance, privacy risks must be continuously reassessed to ensure that anonymization methods do not inadvertently compromise individual confidentiality while still allowing meaningful insights to be drawn.

To mitigate the unintended consequences of privacy-preserving techniques, researchers are exploring more equitable approaches to data protection. Bowen and Snoke (2023) reviewed the prior work on equity in data privacy and introduced an equitable data privacy workflow that integrates fairness considerations into privacy-preserving methods. Majeed and Hwang (2024) proposed an algorithm aimed at balancing privacy protections with the need for equitable data representation, offering a practical approach that highlights ongoing efforts in the field to reconcile these competing priorities.

***Privacy and Confidentiality as a pillar of Data Equity.*** Privacy and confidentiality are fundamental to safeguarding individuals' rights and ensuring that the benefits of data science are equitably distributed across all communities. Protecting sensitive personal information is especially critical for marginalized populations—such as individuals facing exclusion or discrimination based on their health or socioeconomic status—who may face disproportionate harm when their privacy is violated and often have fewer resources to manage those risks (Sannon and Forte, 2022). Effective privacy measures, such as secure data storage and access controls, help to prevent the misuse of information while fostering trust between researchers and communities. Similarly, confidentiality ensures that data shared for research or policy purposes is used responsibly and ethically, preserving the dignity and autonomy of those represented.

However, privacy protections must be carefully balanced to prevent unintended consequences. Overly restrictive privacy measures can limit the availability of high-quality data, obscure disparities and make it more difficult to identify and address structural inequities. Decisions on these trade-offs should involve the communities from which data are collected. Embedding comprehensive privacy and confidentiality safeguards within data equity frameworks helps ensure data remain both ethically managed and sufficiently representative, strengthening the inclusivity and credibility of data-driven decision-making in public health and beyond.

**Other related concepts**

*Justice*

Data Justice emphasizes equitable and fair practices in the collection, use, and governance of data, extending beyond technical considerations to address systemic societal impacts. Taylor (2017) connected data justice to global digital rights, highlighting its role in protecting marginalized groups from the adverse consequences of biased or inequitable data practices. Heeks and Renken (2018) proposed a framework for data justice that emphasizes fairness across data use, handling, and distribution, advocating for practices that align with principles of social justice. Both studies underlined the importance of aligning data practices with broader principles of social justice, ensuring that data-driven systems empower communities rather than exacerbate existing inequalities.

*Trust*

Trust refers to a psychological state characterized by the belief or conviction that another entity will act as expected or promised, even in situations involving uncertainty or risk (Lockey et al., 2021, Rountree et al., 2025b). In the context of AI and ML, trust has become a pivotal concept as the adoption of these technologies grows across various domains. Siau and Wang (2018) defined trust in AI as the extent to which individuals are willing to rely on AI systems, accept their decisions, and collaborate with them by sharing tasks and contributing information. Toreini et al. (2020) emphasized that trustworthiness in AI and ML extends beyond technical reliability to encompass ethical considerations, transparency, and accountability. This broader understanding of trust underscores the need for algorithms that not only perform effectively but also align with societal values, ensuring that users feel confident in their fairness, integrity, and ability to support equitable outcomes. As a foundational element in the human-technology relationship, trust fosters the acceptance and responsible integration of AI and ML into decision-making processes, while mitigating concerns about misuse and unintended consequences.

## 4.2. Data Equity-Related Concepts Rooted in Public Health

Public health has significantly advanced with the integration of modern data science techniques, offering unprecedented opportunities to analyze vast datasets and inform policies aimed at improving health outcomes (Murdoch and Detsky, 2013). By leveraging statistical models and ML methods, researchers have been able to identify disease patterns, predict outbreaks, and optimize resource allocation across domains such as infectious disease control and chronic disease management (Chien et al., 2020, Sedik et al., 2020, Garriga et al., 2022). However, the increasing reliance on data science has also amplified equity challenges, particularly in how data is collected, analyzed, and applied. Observational data, often foundational in public health research, lack the experimental rigor of randomized controlled trials and can introduce biases that distort findings if not carefully addressed (Rothman et al., 2021). Furthermore, the adoption of ML and AI in public health raises ethical concerns, including the risk of perpetuating disparities in care and access for marginalized populations (Obermeyer et al., 2019). These challenges underscore the importance of considering equity at every stage of the public health data life cycle, ensuring that advancements benefit all populations equitably.

Key methodological concepts in public health research — selection bias, representativeness, generalizability, causality, and information bias — play a crucial role in ensuring both the validity and accuracy of research findings. Selection bias arises when the process of recruiting participants leads to systematic differences between the study sample and the target population, potentially distorting associations and limiting the external validity of findings (Rothman et al., 2021). Representativeness assesses whether study results reflect the characteristics of the target population, while generalizability evaluates the applicability of these findings to broader contexts or diverse populations (Gobo, 2004). Causality focuses on identifying the underlying relationships between exposures and outcomes, allowing researchers to design interventions that address root causes rather than superficial correlations. Information bias occurs when errors in measuring exposures, outcomes, or covariates introduce systematic distortions, affecting the accuracy of study conclusions (Alexander et al., 2015a). Together, these concepts underpin efforts to produce equitable research outcomes by identifying and mitigating biases, improving the relevance of findings, and ensuring interventions effectively address health disparities. This section explores these interconnected concepts and their roles in fostering data equity.

**Selection Bias**

***Existing Literature.*** Selection bias broadly refers to distortions that arise from the procedures used to select study subjects or from factors that influence their participation in the study (Porta et al., 2008). It occurs when the relationship between exposure and outcome in the study sample differs from what would be observed in the target population, leading to biased estimates of association or effect (Rothman et al., 2021).

Tripepi et al. (2010) identified five common types of selection bias, including non-response bias, incidence-prevalence bias, loss-to-follow-up bias, confounding by indication bias and volunteer bias. In addition to these, Rojas-Saunero et al. (2023) introduced a complementary framework that focuses on structural mechanisms underlying selection bias, with particular attention to collider-stratification bias and generalizability bias.

To support the formal study of selection mechanisms, Hernán et al. (2004) advocated for the use of directed acyclic graphs (DAGs), offering a visual and conceptual approach to identifying and mitigating selection bias in case-control and cohort designs. Building on this, Lu et al. (2022) proposed a formal causal definition of selection bias as the difference between the true causal effect in the target population and the observed effect among selected individuals. On the risk difference scale, this can be written as:

$$[P(Y(1) = 1) - P(Y(0) = 1)] - [P(Y = 1 \mid A = 1, S = 1) - P(Y = 1 \mid A = 0, S = 1)]$$

Here, $Y(a)$ denotes the potential outcome under exposure level $A = a$, and $S = 1$ indicates that the indivdiual is included in the sample. The term $P(Y(1) = 1) - P(Y(0) = 1)$ represents the population-level average treatment effect, while $P(Y = 1 \mid A = a, S = 1)$ denotes the observed probability of the outcome $Y = 1$ given exposure level $A = a$ and inclusion in the sample. This formulation clarifies that selection bias arises whenever the observed association within the selected sample differs from the population-level causal effect. Lu et al. (2022) further distinguish between type 1 selection bias, which results from conditioning on a collider (e.g., a variable affected by both exposure and outcome), and type 2 selection bias, which stems from conditioning on an effect modifier, leading to differences in causal effects across subgroups.

***Addressing selection bias as a pillar of data equity.*** Selection bias poses a fundamental threat to data equity by systematically excluding or underrepresenting certain groups, resulting in research findings that fail to reflect the diversity of real-world populations. This misrepresentation can ultimately lead to interventions and policies that perpetuate or exacerbate existing inequities.

Beesley and Mukherjee (2022) emphasized that even large-scale health datasets are susceptible to biases introduced through design choices, such as who is included and how variables are measured. These biases do not diminish with increasing sample size; instead, they can produce misleading inferences, inflated type I error rates, and reduced statistical coverage, thereby compromising both the validity and equity of research outcomes.

In practice, selection mechanisms embedded in real-world datasets often limit equitable representation. For example, population-based biobanks like the UK Biobank rely on volunteer participation, which can lead to the overrepresentation of certain demographic groups while underrepresenting others (Fry et al., 2017). In contrast, nationally representative initiatives, such as the NIH All of Us program, use purposeful sampling to oversample underrepresented subgroups, improving diversity. (The All of Us Research Program, 2019). Medical center-based studies tend to recruit patients meeting specific criteria, resulting in an overrepresentation of certain diseases or conditions while excluding others (Pendergrass et al., 2010, Zawistowski et al., 2023). Additionally, nonresponse and consent-related biases can further limit diversity among participants (Kundu et al., 2024).

These forms of selection bias diminish the equity of data by skewing the representation of populations, particularly marginalized groups such as racial and ethnic minorities, rural communities, and individuals with limited healthcare access. Addressing selection bias is therefore essential for producing data-driven insights that are not only scientifically valid but also socially inclusive and just.

**Representativeness**

***Existing Literature.*** The concept of representativeness has been variably defined across the literature. A series of discussions in 2013 (Elwood, 2013, Nohr and Olsen, 2013, Richiardi et al., 2013, Rothman et al., 2013) described representativeness as occurring when the study sample is a simple random sample of the target population. This definition emphasizes equal probability of selection and ensures that the sample mirrors the population's characteristics without systematic deviation (Rudolph et al., 2023).

However, others have argued for a more practical interpretation. Porta et al. (2008) argued that representativeness should not be strictly tied to probabilistic sampling. Instead, they suggested that a study sample can be considered representative if key characteristics or outcomes align with those of the target population, making the concept more practical for observational studies. Expanding on this flexibility, Rudolph et al. (2023) proposed an even broader definition, stating that a sample is representative of a target population if its estimates or interpretations are generalizable to that population. This perspective highlights that representativeness is not solely determined by sampling methods but can also be achieved through thoughtful study design and robust analytical techniques.

To operationalize representativeness, Tipton (2014) introduced a generalizability index $\beta$, which measures the degree of similarity between the sample and population based on the distribution of estimated propensity scores. Specifically, for a set of covariates $X$, let $s = s(X)$ denote the estimated propensity score, and let $f_s(s)$ and $f_p(s)$ be the density functions of $s$ in the sample and population, respectively. Then,

$$\beta = \int \sqrt{f_s(s) f_p(s)}\, ds.$$

The index requires no distributional assumptions and is bounded between 0 and 1, ranging from distinctly unrepresentative to perfectly representative. A value of $\beta \geq 0.90$ is often used as a heuristic threshold for representativeness. Lesko et al. (2017) further contextualized representativeness within the potential outcomes framework, linking it to the projection of causal effect estimates to a target population. They underscored the risks of bias when discrepancies exist between study samples and target populations, particularly regarding effect measure modifiers, and advocated for statistical methods like direct standardization and inverse probability weighting to mitigate these challenges. Building on similar motivations, Dahabreh et al. (2018) proposed identifiability conditions and estimation strategies for generalizing causal inferences from randomized individuals to all trial-eligible individuals. Their approach leverages outcome model-based, inverse probability weighting, and doubly robust estimators to improve validity when extending trial findings beyond trial samples.

***Representativeness as a pillar of data equity.*** Representativeness is a critical component of data equity, as it ensures that study samples reflect the full (social, demographic, genetic and geographical) diversity and needs of the population of focus. It is closely related to selection bias but focuses specifically on how well the sample captures the diversity of the intended population, rather than on the mechanisms that lead to inclusion or exclusion. When datasets disproportionately represent dominant groups, analyses may overlook or even worsen disparities affecting underrepresented communities.

Persistent barriers to inclusion — such as limited digital literacy among older adults, restricted access of incarcerated individuals to consented research process, inaccessibility of surveys to the visually disabled, exclusion of the homeless in household pulse surveys, and underrepresentation of people living in rural areas in health cohorts — continue to limit the participation of these and other groups in data collection and analysis. The challenges of achieving representativeness in research were further exposed during the COVID-19 pandemic (Jagadish et al., 2021). Marginalized communities, such as American Indian/Alaska Native populations, faced unique barriers to research participation, exacerbating systemic inequities and cultural losses (Mays et al., 2022). Similarly, Ponce et al. (2023) described how systemic disparities in areas like healthcare, education, and housing rendered these groups underrepresented in research and policy-making, ultimately compromising equity in public health outcomes and resource allocation.

Importantly, what constitutes adequate representation is context-specific, shaped by local histories, cultural norms, and structures of identity and inclusion. Addressing representativeness requires intentional design choices, inclusive recruitment strategies, and continuous evaluation to ensure that all segments of the target population are appropriately reflected in both data collection and analysis.

**Generalizability**

***Existing Literature.*** Generalizability, also known as external validity, describes the extent to which findings from a study can be extended to other populations, settings, or time periods beyond those directly studied (Borodovsky, 2022). In some literature, this concept is also referred to as transportability (Westreich et al., 2019). While generalizability is closely related to representativeness, Gobo (2004) argued that generalizability involves additional considerations beyond representativeness, such as the trustworthiness of operational definition, the trustworthiness of operationalization, the reliability of method, the suitability of conceptualization, the researchers' accuracy, their degree of success with field relations, and data and interpretation validity. Borodovsky (2022) further distinguished two types of generalizability: generalizations about the state of a complex, non-stationary system at a specific point in time, and generalizations about cause-effect relationships that are partially or entirely invariant across modifying variables such as time, location, or species.

From a causal inference perspective, Pearl and Bareinboim (2014) and Pearl (2015) define generalizability as the capacity to transfer causal effects learned in one population (the source population $\Pi$) to another (the target population $\Pi^*$), under assumptions about their structural differences. Let $A$ repersent a treatment or intervention, and $Y$ denote an outcome of interest. The causal effect of $A$ on $Y$ is expressed using the do-operator, $P(y \mid \mathrm{do}(a))$, which captures the distribution of outcomes that would be observed if all indivdiuals in a population were assigned treatment $a$. The target of generalization is the causal effect in the target population, denoted $P^*(y_a)$, where $y_a$ refers to the potential outcome under treatment $a$ in $\Pi^*$. Assuming that the differences between the source and target populations can be attributed to a set of selection variables $S$, this causal query is reframed as

$$P^*(y_a) = P(y \mid \mathrm{do}(a), S = 1),$$

which defines the causal effect in the subpopulation identified by $S = 1$, that is, the target population $\Pi^*$. The information available for identifying this causal quantity includes experimental estimates from the source population (e.g., $P(y \mid \mathrm{do}(a), x)$, where $X$ is a set of measured covariates), and observational data from both populations (e.g, $P(a, y, x)$ and $P(a, y, x \mid S = 1)$). This identification task is guided by a structural causal model $M$, often represented as a selection diagram, which encodes how the populations differ. Causal transportability holds if the target query $P^*(y_a)$ can be computed using only these inputs under the assumptions encoded in $M$.

An earlier approach by Stuart et al. (2010) framed generalizability as a selection problem and proposed using propensity scores for trial participation to adjust effect estimates, enabling better alignment between study results and target populations. To systematically evaluate generalizability across multiple dimensions, Findley et al. (2021) introduced the M-STOUT framework, a structured approach to evaluating generalizability across six dimensions: Mechanisms, Settings, Treatments, Outcomes, Units, and Time. This framework helps researchers identify potential threats to generalizability and assess when additional assumptions or statistical adjustments are necessary to support valid extrapolation.

A broader synthesis of these developments is provided by Degtiar and Rose (2023), who reviewed generalizability and transportability methods across disciplines. Their work outlines diagnostic strategies, modeling approaches, and estimation techniques, offering both a theoretical foundation and applied guidance for researchers aiming to assess and address external validity bias. They emphasize that external validity requires careful attention not only to covariate overlap and effect heterogeneity, but also to the structure and assumptions underlying different data sources.

Finally, Parikh et al. (2025) demonstrated how insufficient representation in certain covariate regions can prevent reliable causal transport, particularly for subgroups with heterogeneous treatment effects, and further proposed a weighted target average treatment effect (WTATE) to characterize treatment effects in underrepresented populations. Let $w(X) \in \{0, 1\}$ indicate membership in a subgroup of interest, such as a racial or geographic group with limited representation in a randomized controlled trial, and define $\tau_0(X) = E[Y(1) - Y(0) \mid X, S = 0]$ as the conditional treatment effect in the target population. The odds of selection into the trial for a given covariate profile is defined as $\ell(X) = \pi(X)/(1 - \pi(X))$, where $\pi(X) = P(S = 1 \mid X)$. Then, the WTATE for subgroup $w$ is given by

$$\tau_0^w = \frac{\pi_w}{1 - \pi_w} \cdot \frac{E\left[\frac{w(X)\tau_0(X)}{\ell(X)} \mid S = 1\right]}{E[w(X) \mid S = 1]},$$

where $\pi_w = E[w(X)S]$ represents the proportion of individuals from subgroup $w(X) = 1$ who are included in the trial. This formulation expresses how treatment effects in underrepresented groups can be identified by reweighting trial data, using inverse odds of selection to correct for lack of representation. Unlike generalizability indices that focus solely on sample selection probabilities, the WTATE approach incorporates treatment effect heterogeneity, offering a more nuanced tool for identifying and addressing generalizability gaps in causal inference.

Together, these perspectives highlight that generalizability is not solely a matter of sampling but also of conceptual and methodological rigor. Ensuring that findings can meaningfully inform decision-making in real-world contexts is essential for both scientific validity and equitable public health outcomes.

***Generalizability as a pillar of data equity.*** Generalizability plays a crucial role in data equity by ensuring that findings from a research study are valid across varied populations and settings of interest. When generalizability is compromised, the resulting policies and interventions risk perpetuating inequities by benefiting only those groups well-represented in the original study while neglecting the needs of others.

Yom et al. (2022) emphasized the importance of generalizability in public health research, arguing that limitations in generalizability or reproducibility may result from mechanisms that operate differently across populations or from structural limitations in delivering therapies appropriately. These concerns are particularly relevant in studies relying on electronic health records (EHRs), where biases introduced by incomplete data, poor data quality, and the overrepresentation of structurally privileged populations can further hinder generalizability (Boyd et al., 2023a, Boyd et al., 2023b). As a result, insights derived from such datasets may be poorly suited to inform interventions in low-resource or underserved settings.

Singh et al. (2022) further demonstrated the risks of limited generalizability through an evaluation of mortality risk prediction models. They found substantial variability in model performance when applied across different countries, hospitals, or time periods — raising concerns about the reliability of models developed in data-rich environments when transferred to low-resource contexts. Without careful consideration of generalizability, data-driven tools and policies may unintentionally exacerbate inequities rather than mitigate them.

**Causality**

***Existing Literature.*** Causality in public health focuses on uncovering the relationships between exposures, interventions, and health outcomes, which is critical for designing effective strategies that address the root causes of disease and health inequities. Unlike simple associations, causal relationships provide insights into the mechanisms that drive health outcomes, enabling the development of interventions that are not only effective but also equitable (Rothman and Greenland, 2005, Glass et al., 2013). These insights form the basis of evidence-based public health, allowing actions to be targeted, justified, and impactful.

Modern causal inference frameworks, including structural causal models (SCMs) and directed acyclic graphs (DAGs), have become indispensable tools in epidemiology and public health. These tools help represent causal relationships explicitly, clarify assumptions, identify potential confounding pathways, and assess the conditions under which valid causal estimates can be obtained (Pearl et al., 2016).

Although randomized controlled trials are the gold standard for establishing causality, their application is frequently limited by ethical, logistical, or temporal constraints (Deaton and Cartwright, 2018, Rothman et al., 2021). Observational studies thus become the primary source of data for causal inference in public health, relying on statistical techniques and careful study design to emulate randomized experiments (Glass et al., 2013). However, confounding bias poses a significant challenge in observational studies, distorting the association between exposures and outcomes due to the influence of extraneous factors, often referred to as confounders (Alexander et al., 2015a). Formally, confounding can be defined in the potential outcomes framework: suppose $A$ is an exposure, $Y$ is an outcome, and $X$ is a covariate. Then $X$ is said to be a confounder if it is associated with both the exposure and the potential outcomes. In this case, unless appropriate adjustment for $X$ is made, the observed association between exposure levels $E[Y \mid A = a] - E[Y \mid A = a^*]$, does not equal the causal effect, $E[Y(a)] - E[Y(a^*)]$. For instance, in a study predicting patient readmission rates, socioeconomic status may act as a confounder. Limited access to healthcare resources among individuals with lower socioeconomic status can lead to poorer medical conditions, which influence both the input variables (e.g., health status) and the model predictions, thereby skewing results.

To address these challenges, various causal inference methods have been developed for public health research. He et al. (2016) reviewed approaches such as propensity score matching and marginal structural models, which are particularly valuable for mitigating confounding in observational studies where randomization is infeasible. These tools enhance the credibility of causal findings, enabling researchers to derive actionable insights that can guide equitable and effective public health policies and interventions.

***Causality as a pillar of data equity.*** Causality is essential to data equity because it enables researchers and policymakers to identify and intervene on the structural and systemic drivers of variations in health outcomes rather than surface-level associations. Health inequities often arise from deeply rooted social determinants, such as poverty, discrimination, housing instability, and unequal access to healthcare and education (Marmot et al., 2008, Braveman et al., 2011). Without a robust causal framework, data-driven interventions risk being reactive and superficial — for example, focusing on individual behaviors such as smoking or diet without addressing the underlying social conditions, such as poverty, limited education, or identity-driven systemic biases, that shape those choices.

Understanding causal relationships allows researchers to distinguish between confounding variables and true mechanisms of disparity, improving the precision and fairness of interventions (Pearl et al., 2016). For example, observational studies that lack proper adjustment for socioeconomic status or structural racism may misattribute disparities to individual behaviors rather than systemic barriers, leading to inequitable or ineffective policy recommendations (Boyd et al., 2020). To prevent this, causal inference methods — such as propensity score methods, instrumental variables, and marginal structural models — provide the essential tools to emulate randomized trials using real-world data, a critical capability in public health contexts where experimentation is often infeasible (Glass et al., 2013). By embedding causality into data equity frameworks, researchers can better ensure that interventions are not only evidence-based but also ethically grounded and, critically, address the root causes of disparities rather than just their manifestations.

**Information Bias**

***Existing Literature.*** Information bias refers to systematic errors in the measurement or recording of key study variables, leading to distorted estimates of association and potentially invalid conclusions (Alexander et al., 2015b). It arises when exposure, outcome, or confounding variables are not measured or recorded correctly, leading to errors in data interpretation and potentially undermining the validity of research findings (Kesmodel, 2018).

A common form of information bias is misclassification bias, which occurs when individuals are incorrectly categorized with respect to their exposure or outcome status. This can be non-differential, where errors occur equally across groups and typically attenuate associations, or differential, where errors vary by exposure or outcome status and can bias estimates in unpredictable directions (Tripepi et al., 2010).

Recall and reporting bias are also frequent in public health research, especially when relying on self-reported data. For instance, individuals with a disease may be more likely to recall or report exposures, while exposed individuals might be more attentive to symptoms or health events than unexposed counterparts. These inconsistencies in data quality can skew results, particularly in case-control or survey-based studies (Alexander et al., 2015b).

In addition, measurement error — caused by instrument inaccuracy, environmental variation, or subjective reporting — can introduce both random noise and systematic bias (Althubaiti, 2016). For example, a miscalibrated device may consistently under- or overestimate physiological measures, distorting associations and leading to flawed conclusions. From a causal perspective, measurement error arises when the variables used in the analysis (e.g., exposure $A^*$ and outcome $Y^*$) do not correspond to their true counterparts $A$ and $Y$, due to imperfect measurement. This discrepancy can bias the estimated associations even in the absence of confounding or selection bias. Generally, the observed associational risk ratio $P(Y^* = 1 \mid A^* = 1)/P(Y^* = 1 \mid A^* = 0)$ will differ from the causal risk ratio $P(Y(1) = 1)/P(Y(0) = 1)$. In such settings, the usual identifiability assumptions — exchangeability, positivity, and consistency — are violated, and causal effects cannot be recovered from the observed data alone (Hernán and Robins, 2010).

Missing data is another critical source of information bias, particularly when data are missing not at random (Little and Rubin, 2019). This issue directly threatens data equity, as marginalized populations or those with limited healthcare access are more likely to be excluded due to incomplete records. Consequently, their experiences may be excluded from analyses, potentially obscuring the very health disparities researchers aim to study and reinforcing existing inequities.

***Addressing information bias as a pillar of data equity.*** As shown above, information bias poses a fundamental threat to data equity. When key variables are mismeasured, misclassified, or missing, the resulting inaccuracies often obscure the true nature of health disparities. Addressing information bias is therefore critical to data equity, as it ensures that study variables are measured accurately and consistently regardless of the population from which it comes.

In the absence of perfectly measured data and inevitable loss to follow-up, proper calibration and imputation methods can improve the quality of inference. This includes developing standardized protocols, thoroughly training interviewers and technicians, conducting pilot studies, and considering the likelihood of misclassification to assess the direction of potential bias (Alexander et al., 2015b). By embedding these rigorous measurement and data collection strategies into study designs, researchers can produce findings that more accurately serve all communities, ultimately ensuring that data science promotes fairness, inclusivity, and justice.

### 4.3. Computable Metrics for Data Equity-Related Concepts

In earlier sections, we explored key concepts of data equity as they are understood in both computer science and public health. This section synthesizes those discussions by demonstrating how each concept can be translated into practice using measurable evaluation metrics and implementation tools.

Table 1 provides a structured summary of these data equity principles, linking each to quantitative metrics and practical methods that support assessment, monitoring, and mitigation. By highlighting these connections, we aim to bridge conceptual understanding with actionable strategies for advancing equity in data-driven systems.

Table 1 Ten core concepts drawn from computer science and public health that underpin data equity. The blue rows reflect concepts commonly emphasized in computer science, and the green rows represent public health priorities. Each concept is accompanied by a working definition and examples of quantitative metrics or methods used to assess or operationalize it in practice.

| Conceptual Pillars | Definition | Evaluation Metrics/Methods (examples) | Tools (examples) |
|---|---|---|---|
| Fairness | The principle of ensuring equal treatment regardless of differences in protected attributes. | Equalized Odds, Equal opportunity, Demographic Parity | Fairlearn, AI Fairness 360 |
| Accountability | A set of preventative or corrective measures that hold the creators, designers, and users of data science and AI systems responsible for the outcomes their algorithms produce. | Incident Response Time, Audit Compliance Rate | Algorithmic Impact Assessments, Little Sis Database |
| Transparency | The degree to which an algorithm's functioning, decision-making process, and outcomes are clear, interpretable, and accessible to users and interest-holders. | Feature Importance Spread, Alpha-Feature Importance, Quantitative Input Influence | InterpretML, Holistic AI library, LIME, SHAP |
| Ethics | The principles of right and acceptable conduct, rooted in societal norms and values, and involves reflective reasoning when such values and norms clash or are unclear. | Ethical Composite Scores | The Ethical Toolkit for Engineering/Design Practice, AI Explainability 360 |
| Privacy | The set of protections individuals or populations have for their own information | Re-identification Risk, Privacy Loss | PySyft, Secure Multi-Party Computation |
| Confidentiality | The obligation of entities handling personal information to prevent unauthorized disclosure. | Confidentiality Metrics | k-anonymity, Differential Privacy, Access Control Mechanisms |
| Selection Bias | A systematic error in the findings due to observed or unobserved factors affecting selection or exclusion of units into a study sample. | Standardized Mean Differences | Propensity Score Matching, Inverse Probability Weighting, Heckman Correction |
| Representativeness | The extent to which study samples reflect the socio-demographic diveristy and needs of the target population. | Participation-to-prevalence Ratio, Generalizability Index | Stratified Sampling, Weighting |
| Generalizability | The extent to which the findings of a study are valid across the diverse target populations and settings of interest. | Generalization Error | Meta-Analysis, External Dataset Validation |
| Confounding Bias (Causality) | The ability to identify causal relationships between exposures and outcomes, after removing confounding bias and satisfying conditions for identifiability and estimability with observed data. | Sensitivity Analysis, E-value | Propensity Score Matching, Instrumental Variables |
| Information Bias | A distortion in the measure of association caused by a lack of accurate measurements of key study variables. | Sensitivity Analysis, Cohen's Kappa, Intraclass Correlation Coefficient | Multiple Imputation, Regression Calibration |

Fairness can be quantitatively assessed using a variety of metrics that evaluate whether model predictions are consistent across different values of sensitive attributes (Rountree et al., 2025a). Equalized odds and equal opportunity measure whether true positive and/or false positive rates are similar across groups. Demographic parity evaluates whether the predicted outcome is independent of sensitive features, while predictive parity assesses whether the precision (positive predictive value) is equal across groups. Additional metrics such as predictive equality and false negative rate parity examine disparities in false positive and false negative rates, respectively. These metrics provide concrete, computable criteria for identifying disparities in model performance and are widely used in fairness audits. To support implementation, several open-source toolkits offer extensive functionality for fairness assessment and mitigation. AI Fairness 360 (Bellamy et al., 2018), available in both Python and R, and Fairlearn (Weerts et al., 2023), designed for Python, include numerous metrics and support pre- and post-processing mitigation strategies. The fairness R package (Kozodoi and V. Varga, 2021) provides a rich suite of tools for computing and comparing fairness metrics, supporting model diagnostics and transparency efforts in statistical analysis.

Accountability in data systems refers to the obligation of organizations to ensure their AI systems operate transparently, ethically, and in compliance with established regulations. While there is no universally accepted computable metric for algorithmic accountability, operational indicators such as incident response time and audit compliance rate have been used in practice to reflect the timeliness and thoroughness of accountability processes (Kavishwar, 2025). Tools like Algorithmic Impact Assessments (AIAs) (Metcalf et al., 2021) provide structured frameworks for evaluating potential societal impacts of AI systems prior to deployment, promoting proactive risk management and engagement of involved parties. Additionally, resources such as the LittleSis Database (The Public Accountability Initiative, 2025) offer platforms for mapping relationships between influential entities, enhancing transparency by revealing connections that may affect AI governance and decision-making processes.

Transparency in ML systems can be evaluated using quantitative explainability metrics that assess how clearly model outputs can be understood and interpreted. Two informative metrics are the feature importance spread and alpha-feature importance, proposed by Munoz et al. (2024). The feature importance spread measures how concentrated the explanatory power is among the features — higher concentration suggests that only a few features dominate, making interpretations more straightforward. The alpha-feature importance quantifies the minimum proportion of features needed to account for a specified threshold (e.g., 90%) of the total feature importance, offering a concise representation of the most influential variables. These metrics enable a systematic assessment of model interpretability and help practitioners evaluate whether model explanations are accessible and meaningful across different user groups. To complement these metrics, several tools have been developed to generate interpretable explanations of model behavior, such as LIME (Ribeiro et al., 2016) , which approximates model behavior locally by fitting interpretable models around individual predictions, and SHAP (Lundberg and Lee, 2017) , which uses game-theoretic Shapley values to attribute feature contributions for both global and local explanations. Model-agnostic platforms such as InterpretML (Nori et al., 2019)  and Holistic AI's explainability suite (Holistic AI, 2024) offer interactive visualizations and model inspection functionalities. These resources enable deeper scrutiny of model behavior and support more transparent, trustworthy deployment in practice.

Ethical considerations in AI systems are increasingly being formalized through composite assessment frameworks that unify multiple dimensions of responsible AI design. One such framework, proposed by Dwivedi et al. (2024), identifies and quantifies a range of ethical indicators — including fairness, accountability, openness, and bias mitigation — by aggregating them into composite scores. These scores provide a structured and interpretable summary of an AI system's ethical standing, allowing for comparison across models and supporting more standardized ethical auditing. Building on this formalization, practical toolkits have emerged to help operationalize ethical reflection and embed ethical values into technology workflows. The Ethical Toolkit for Engineering/Design Practice from Santa Clara University offers structured methods for integrating ethical analysis into day-to-day engineering and design tasks, aiming to cultivate habits of ethical judgment that are both explicit and repeatable (Vallor et al., 2018). Complementing this, AI Explainability 360 (Arya et al., 2019) provides a diverse set of explainability algorithms and evaluation metrics that facilitate ethical assessments by enhancing model transparency. Together, these resources support practitioners in proactively identifying ethical risks, aligning technical development with human values, and promoting responsible AI adoption across diverse application contexts.

Privacy and confidentiality are central to data equity, particularly in contexts involving sensitive personal information. Privacy can be quantitatively assessed using metrics such as re-identification risk (Benitez and Malin, 2010, Carey et al., 2023), which estimates the likelihood that anonymized data can be linked back to an individual, and privacy loss (Sommer et al., 2018), which quantifies the informational leakage from data output, often formalized through differential privacy frameworks. Tools such as PySyft (Ziller et al., 2021) and secure multi-party computation protocols (Du and Atallah, 2001) enable privacy-preserving data analysis by allowing computations on encrypted or decentralized data without revealing individual-level information.

Confidentiality, distinct from privacy, emphasizes the protection of specific data attributes from unauthorized disclosure, even when records cannot be directly linked to individuals. Domingo-Ferrer et al. (2021) introduced a set of general confidentiality metrics grounded in the permutation model, which conceptualizes anonymization as a combination of permutation and noise addition. These metrics facilitate quantitative comparison of privacy-preserving approaches, including k-anonymity (Sweeney, 2002) and ε-differential privacy (Dwork et al., 2006). In practice, access control mechanisms (Samarati and De Vimercati, 2000) play a critical role in supporting confidentiality by enforcing data access policies and limiting exposure to sensitive information based on user roles and permissions.

Selection bias undermines data equity by distorting the association between exposures and outcomes due to non-random inclusion of individuals in the dataset. A common way to quantify selection bias is through standardized mean differences, which measure the imbalance in covariate distributions between the included sample and the target population (Steiner et al., 2011). To mitigate selection bias, several methods have been developed. Propensity score matching aligns treated and untreated groups on observed covariates to approximate randomized comparisons (Bai, 2011). Inverse probability weighting adjusts for the probability of inclusion in the sample, reweighting individuals to restore population-level balance (Beesley and Mukherjee, 2022). Heckman correction models are particularly useful when selection depends on unobserved variables, using a two-step procedure to correct for endogenous selection effects (Heckman, 1976). Together, these tools provide a principled approach for diagnosing and reducing selection bias in both observational and algorithmic settings.

Representativeness can be assessed using computable metrics such as the participation-to-prevalence ratio (PPR), which compares the proportion of individuals from a demographic group enrolled in a study to their proportion in the broader population. A PPR between 0.8 and 1.2 is often interpreted as indicating adequate representation (Chen et al., 2022). To improve representativeness, researchers can implement stratified sampling (Neyman, 1992) and apply sampling weights (Barratt et al., 2021) to account for disparities in subgroup inclusion, ultimately producing findings that are more applicable and equitable across diverse populations. However, these approaches assume that proportional representation along observed characteristics is sufficient for supporting valid inference. Parikh et al. (2025) challenged this view, arguing that such metrics may overemphasize imbalances in features unrelated to treatment effect heterogeneity, while overlooking covariate regions where estimation is unreliable due to sparse data support. Tipton (2014) offered a complementary perspective by proposing a generalizability index based on the overlap of propensity score distributions between the study and target populations. This approach shifts the focus from marginal subgroup counts to functional coverage of the covariate space relevant for inference.

Generalizability refers to the extent to which findings from a study can be validly applied to other populations, settings, or timeframes beyond the original study context (Borodovsky, 2022). It can be assessed using computable metrics such as the generalization error (Jakubovitz et al., 2019), which quantifies the discrepancy between predicted outcomes in the source and target populations, and a recently proposed external validity measure (Bo and Galiani, 2021), which evaluates how frequently the original causal conclusion of a study remains valid across reweighted samples representing new populations. Methods to improve generalizability include meta-analysis (Degtiar and Rose, 2023), which synthesizes findings across studies to identify consistent effects, and external dataset validation (König et al., 2007), which tests model performance in new populations. Together, these tools help ensure that data-driven insights retain relevance and fairness across diverse real-world contexts.

Causal validity can be compromised by confounding bias, which arises when the association between exposure and outcome is distorted by extraneous variables. To evaluate the robustness of causal claims against unmeasured confounding, sensitivity analysis techniques such as the E-value offer a computable metric that quantifies the minimum strength of association an unmeasured confounder would need to nullify the observed effect (VanderWeele and Ding, 2017, Haneuse et al., 2019). Methodological approaches like propensity score matching (Austin, 2011) and instrumental variable analysis (Zhang et al., 2018) are commonly used to address measured and unmeasured confounding, respectively. These tools are widely implemented in statistical software, facilitating the application of causal inference frameworks in both observational and experimental settings.

Information bias occurs when key variables such as exposures, outcomes, or covariates are inaccurately measured or classified, leading to distorted effect estimates (Alexander et al., 2015b). Several computable metrics can be used to assess the extent of measurement error and misclassification. Cohen's kappa is a metric used to evaluate the agreement between two categorical data sources, correcting for agreement that might occur by chance. (McHugh, 2012). For continuous variables, the intraclass correlation coefficient (ICC) assesses the reliability or consistency of repeated measurements (Kim, 2013). Additionally, sensitivity analysis techniques can also be applied to evaluate how varying levels of measurement error influence the robustness of study conclusions (VanderWeele and Li, 2019). To address information bias, researchers often employ methods such as multiple imputation to handle missing or mismeasured values (Cole et al., 2006) and regression calibration to correct bias in the presence of measurement error (Spiegelman et al., 1997). These methods help improve data accuracy and the validity of downstream statistical inferences.

## 5. Conclusion and Recommendations

This paper emphasizes the vital role of data equity as a guiding principle for public health data science. Real-world examples, from the inequity in EHR and GWAS research to the racial disparities in pulse-oximetry studies, illustrate the tangible consequences of neglecting equity in data practices. These illustrative examples highlight the need of adopting more inclusive and equitable data strategies to prevent harm and better serve populations that are unseen in our current data systems.

To address these challenges, we take an interdisciplinary approach that integrates insights from computer science and public health to propose a structured framework for evaluating and advancing public health data science and data equity. While data equity remains an emerging and evolving concept, its implementation is essential for fostering inclusivity and fairness in both research and decision-making. We discuss a set of foundational principles, including fairness, accountability, transparency, ethics, privacy and confidentiality, selection bias, representativeness, generalizability, causality, and information bias, which together serve as pillars for equitable data systems.

To operationalize data equity, we suggest a self-auditing process for data life cycle (Figure 4) that outlines key principles to be considered at each phase.

In the **study and sampling design phase**, equity begins with inclusive representation and thoughtful protocol development. Fairness calls for the intentional inclusion of minority groups, such as recruiting through trusted community organizations, designing eligibility criteria that reflect real-world conditions, and involving community members in planning and recruitment. Transparency fosters trust through open communication about research goals and selection criteria. Ethical design safeguards participants from harm, while efforts to mitigate selection bias reduce systematic exclusion. Representativeness ensures that samples reflect the diversity of the target population, and attention to generalizability allows findings to extend beyond the immediate study context.

During **data collection,** fairness is reflected in sampling practices that not only follow representative designs but also actively reach and retain hard-to-reach groups. Transparency requires clear communication of data collection methods and objectives, fostering trust and accountability. Ethics ensure that data are collected responsibly, upholding participants' rights and avoiding harm. Privacy and confidentiality are paramount to safeguarding sensitive information, particularly in public health research, where data often includes personal or medical details. Ensuring representativeness at this phase means capturing the diversity of the target population, reducing biases that may arise from over- or under-sampling specific groups. Lastly, addressing information bias through rigorous standardization and validation protocols ensures that collected data are accurate and reliable for downstream analyses.

In the **analysis phase,** fairness guides the use of equitable modeling techniques and statistical methods that address biases and avoid disadvantaging underrepresented populations. Transparency is achieved through detailed documentation of analytical decisions, such as model selection and methodological choices, allowing for accountability and reproducibility. Ethical principles shape the interpretation of results to ensure equity and mitigate harm. Researchers must address selection bias and representativeness to avoid skewed or inequitable conclusions, while mitigating information bias, such as misclassification or measurement errors, strengthens the validity of findings. Additionally, integrating causality into the analysis phase helps identify and account for confounding variables, enabling robust and actionable conclusions about the relationships between exposures and outcomes.

When **interpreting results**, fairness ensures that conclusions are inclusive and reflective of diverse populations, avoiding the reinforcement of existing inequities. Transparency in this phase entails clearly communicating the limitations, assumptions, and the context in which results should be interpreted. Ethical considerations guide researchers to acknowledge potential biases and frame conclusions responsibly. Accountability involves taking ownership of the interpretations and, importantly, communicating results back to the communities from which the data originated. Incorporating causality allows researchers to differentiate between mere associations and actionable relationships, ensuring interpretations support effective interventions. Finally, generalizability ensures that insights are applicable to broader populations or settings, addressing the variability and needs of diverse communities.

Finally, the **translation** phase focuses on applying research findings to real-world policies, interventions, or tools, ensuring that the benefits are equitably distributed across communities. Fairness ensures that solutions meet the needs of all groups, particularly those historically underserved. Transparency involves clear communication of how findings were derived and their implications for practice, enabling relevant parties to make informed decisions. Ethical use of data aligns implementation with broader social values, while accountability underscores the shared responsibility of researchers, policymakers, and other actors in ensuring equitable application. Causality ensures that translated interventions are based on sound causal relationships, maximizing their effectiveness.

Lastly, generalizability guarantees that these solutions are both relevant and effective for the diverse populations they are intended to serve — supporting population-level impact and long-term trust.

Together, this framework promotes a shift from passive awareness of data inequities to proactive and measurable practices that center equity throughout the entire data science process. By embedding these principles across the life cycle of public health data — from design to action — we lay the groundwork for more just, inclusive, and effective systems that serve the needs of all communities.

However, it is important to recognize that data equity alone does not guarantee learning/information equity or decision equity. How much data do we actually need to achieve equal predictability? That may not be just dictated by the sample size and representation, but involve sophisticated prediction power calculations (Riley et al., 2020). Ensuring that data are equitably collected, analyzed, predicted, and interpreted is only one part of the broader ecosystem; true public health equity also depends on how information is communicated and how public health decisions are made. Information and knowledge alone do not empower an individual or a population with agency to take decisions around their health. Systemic changes are needed to promote decision equity. As such, advancing data equity must be accompanied by parallel efforts in information theory and structural changes to empower individuals and communities with informed decision-making around their and others' health.

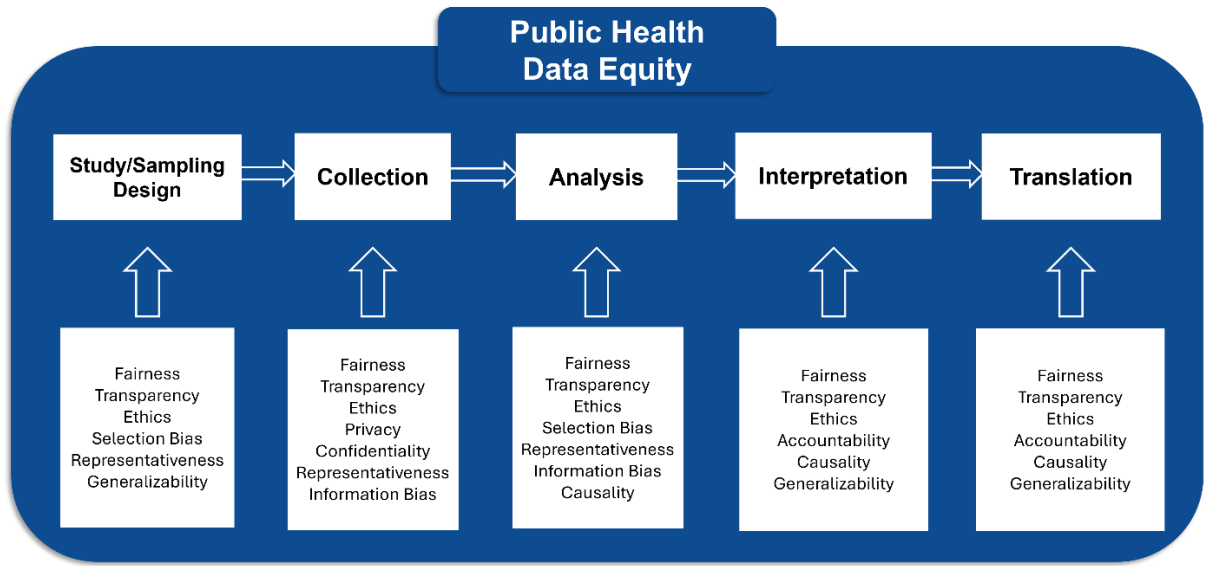

**Figure 4** Self-auditing process for data life cycle. This framework illustrates how different key factors are integrated across the stages of the data life cycle: study/sampling design, collection, analysis, interpretation, and translation. Each stage highlights specific challenges and ethical considerations, ensuring data equity throughout the process.

Gobo, G. (2004). Sampling, Representativeness and Generalizability. <u>Qualitative Research Practice</u>. 1 Oliver's Yard, 55 City Road, London England EC1Y 1SP United Kingdom, SAGE Publications Ltd**:** 405-426.

Gottlieb, E. R., J. Ziegler, K. Morley, B. Rush and L. A. Celi (2022). "Assessment of Racial and Ethnic Differences in Oxygen Supplementation Among Patients in the Intensive Care Unit." <u>JAMA Internal Medicine</u> **182**(8): 849.

Guidotti, T. L. (2002). "Apportionment in Asbestos-Related Disease for Purposes of Compensation." <u>INDUSTRIAL HEALTH</u> **40**(4): 295-311.

Hadi, M. U., Q. A. Tashi, A. Shah, R. Qureshi, A. Muneer, M. Irfan, A. Zafar, M. B. Shaikh, N. Akhtar, J. Wu, S. Mirjalili and M. Shah (2024). Large Language Models: A Comprehensive Survey of its Applications, Challenges, Limitations, and Future Prospects, Preprints.

Hagendorff, T. (2020). "The Ethics of AI Ethics: An Evaluation of Guidelines." <u>Minds and Machines</u> **30**(1): 99-120.

Haneuse, S., T. J. VanderWeele and D. Arterburn (2019). "Using the E-Value to Assess the Potential Effect of Unmeasured Confounding in Observational Studies." <u>JAMA</u> **321**(6): 602-603.

Hardt, M., E. Price and N. Srebro (2016). <u>Equality of Opportunity in Supervised Learning</u>. NIPS '16: The 30th International Conference on Neural Information Processing Systems, Barcelona, Spain, Curran Associates Inc.

Harishbhai Tilala, M., P. Kumar Chenchala, A. Choppadandi, J. Kaur, S. Naguri, R. Saoji and B. Devaguptapu (2024). "Ethical Considerations in the Use of Artificial Intelligence and Machine Learning in Health Care: A Comprehensive Review." <u>Cureus</u> **16**(6): e62443.

Harman, L. B., C. A. Flite and K. Bond (2012). "Electronic health records: privacy, confidentiality, and security." <u>AMA journal of ethics</u> **14**(9): 712-719.

He, H., P. Wu and D.-G. Chen (2016). <u>Statistical causal inferences and their applications in public health research</u>, Springer.

Heckman, J. J. (1976). The common structure of statistical models of truncation, sample selection and limited dependent variables and a simple estimator for such models. <u>Annals of economic and social measurement, volume 5, number 4</u>, NBER**:** 475-492.

Heeks, R. and J. Renken (2018). "Data justice for development: What would it mean?" <u>Information Development</u> **34**(1): 90-102.

Henry, N. R., A. C. Hanson, P. J. Schulte, N. S. Warner, M. N. Manento, T. J. Weister and M. A. Warner (2022). "Disparities in Hypoxemia Detection by Pulse Oximetry
78

**Glossary**

Data equity: The shared responsibility for fair data practices that respect and promote human rights, opportunity and dignity. It requires strategic, participative, inclusive, and proactive collective and coordinated action to create a world where data-based systems promote fair, just and beneficial outcomes for all individuals, groups and communities. It recognizes that data practices – including collection, curation, processing, retention, analysis, stewardship and responsible application of resulting insights – significantly impact human rights and the resulting access to social, economic, natural and cultural resources and opportunities (World Economic Forum, 2024).

Fairness: The principle of ensuring equal treatment regardless of differences in protected attributes (Jagadish et al., 2022).

Accountability: A set of preventative or corrective measures that hold the creators, designers, and users of AI systems responsible for the outcomes their algorithms produce (Memarian and Doleck, 2023).

Transparency: The clarity and openness with which an algorithm's functioning, decision-making processes, and outcomes are made understandable and accessible to relevant partners, such as patients and healthcare providers (Memarian and Doleck, 2023).

Ethics: The principles of right and acceptable conduct, rooted in societal norms and values, and involves reflective reasoning when such values and norms clash or are unclear (Stahl et al., 2016).

Privacy: A set of protections for information that are held by the individual whose information it is (Hulkower et al., 2020).

Confidentiality: A duty held by the person who receives a patient's health information to prevent unauthorized disclosure (Hulkower et al., 2020).

Trust: A psychological state comprising the intention to accept vulnerability based upon positive expectations of the intentions or behaviour of another entity (Lockey et al., 2021).

Justice: The principle of fairness in how people are made visible, represented, and treated as a result of their production of digital data (Taylor, 2017).

Selection Bias: Any deviation between the parameter of interest in the target population and the expected value of the estimate in the sample, caused by the processes through which observations are included in the sample (Rojas-Saunero et al., 2023).

Representativeness: The extent to which a study sample's results are generalizable to a well-defined target population (Rudolph et al., 2023).

Generalizability: The extent to which the findings of a study can be applied to populations, settings, or timeframes beyond the specific context in which the study was conducted (Borodovsky, 2022).

Causality: The property of being causal, the presence of cause, or ideas about the nature of the relations of cause and effect (Susser, 2001).

Information Bias: A distortion in the measure of association caused by a lack of accurate measurements of key study variables (Alexander et al., 2015b).

**Supplementary Materials for "A Framework for Defining and Enhancing Data Equity in Public Health Data Science"**

Yiran Wang, PhD[1], Alicia E Boyd, PhD[1], Lillian Rountree, MS[1], Yi Ren, MS[1], Kate Nyhan, MLS[2,3], Ruchit Nagar, MD[4,5], Jackson Higginbottom, MPH[1], Megan L Ranney, MD[6], Harsh Parikh, PhD[1,7], Bhramar Mukherjee, PhD[1,8]

[1] Department of Biostatistics, Yale School of Public Health, Yale University, 60 College Street, New Haven, CT 06510, USA
[2] Harvey Cushing/John Hay Whitney Medical Library, Yale University, 333 Cedar St, New Haven, CT 06510, USA
[3] Department of Environmental Health Sciences, Yale School of Public Health, Yale University, 60 College Street, New Haven, CT 06510, USA
[4] Department of Pediatrics, Yale – New Haven Hospital, New Haven, CT 06510, USA
[5] Department of Internal Medicine, Yale – New Haven Hospital, New Haven, CT 06410, USA
[6] Department of Health Policy and Management, Yale School of Public Health, Yale University, 60 College Street, New Haven, CT 06510, USA
7 Department of Biostatistics, Johns Hopkins Bloomberg School of Public Health, Baltimore, MD 21205, USA
[8] Department of Chronic Disease Epidemiology, Yale University, New Haven, CT 06510, USA

Corresponding author:
Bhramar Mukherjee
60 College St
New Haven, CT 06510
bhramar.mukerhjee@yale.edu
(203) 737-8644

**S1. Supporting Information for EHR Publication Analysis in Chapter 2.1**

This section provides a detailed description of the methodology used to analyze disparities in geographic representation of Electronic Health Records (EHR)-related research, as presented in Chapter 2.1 of the main document. The supplementary methods outlined below ensure transparency and reproducibility of the study findings.

To assess the volume and distribution of EHR-related research across geographic populations, we conducted a systematic literature search using Ovid, a widely used research platform that provides access to multiple biomedical and life sciences databases. In addition to Embase, described in the main text, we included MEDLINE to broaden the scope of analysis, given its extensive coverage of peer-reviewed publications in healthcare, public health, and medical informatics. While MEDLINE indexes publications from 1946 onward and Embase from 1974 onward, this study focuses on EHR-related research published between 2014 and 2024. All records were retrieved on February 6, 2025, ensuring a consistent dataset for analysis.

We employed two complementary search strategies: a keyword-based search using the multi-purpose (.mp.) operator, and a structured search using MeSH (in MEDLINE) or Emtree (in Embase). While these methods are summarized in the main text, this section provides additional details regarding their implementation. The .mp. search retrieves studies by scanning multiple fields to ensure broad coverage of terminology variations. However, the specific fields it searches differ across databases. In MEDLINE, it includes fields such as the title, abstract, original title, subject heading word, protocol supplementary concept word, and unique identifier, whereas in Embase, it extends to additional metadata fields such as drug trade names, device manufacturers, and keyword indexing. The search was performed using the query "(EHR OR EMR OR electronic health record* OR electronic medical record* OR digital medical record* OR electronic patient record*).mp. ", ensuring that studies discussing EHR across different contexts were included, even if they had not been assigned standardized subject headings.

For the structured search method, we retrieved publications based on pre-assigned subject categories. MeSH, developed by the National Library of Medicine (NLM), is a hierarchical thesaurus used for indexing and retrieving biomedical and health-related information across MEDLINE, PubMed, and other NLM databases. It organizes concepts into a structured framework, enabling precise and consistent categorization of research topics. Similarly, Emtree serves as the controlled vocabulary for Embase, covering a broad range of biomedical and life sciences terminology, including drugs, diseases, medical devices, and key life science concepts. It facilitates deep, full-text indexing of articles, enhancing the discoverability of biomedical research. To identify EHR-related studies, we applied the search queries "exp Electronic Health Records/" in MEDLINE and "exp Electronic Health Record/ OR Electronic Medical Record/ OR Electronic Patient Record/" in Embase. This approach ensured the inclusion of all publications explicitly indexed under these terms, providing a structured and comprehensive retrieval of relevant studies. However, it may overlook studies that discuss EHR without being formally indexed under these subject headings. A detailed manual introducing provides guidance on using the Ovid platform to retrieve publications related to EHR can be found on [Github](Github).

Applying these search strategies, we identified 19,490 geo-tagged publications from MEDLINE and 37,729 from Embase using the .mp. search, while the

MeSH/Emtree-based search retrieved 9,534 geo-tagged publications from MEDLINE and 31,489 from Embase. Geo-tagging was applied to ensure that only studies with assigned geographic indexing terms were included. To ensure consistency in geographic classification, studies were segmented using database-specific geographic indexing terms, with MeSH applied in MEDLINE and Emtree in Embase.

While the general approach remained the same, specific terms varied between databases. For example, studies from the United States were retrieved using the query "exp United States/ OR Puerto Rico/ OR United States Virgin Islands/" in MEDLINE, whereas in Embase, the corresponding query was "exp United States/." Similarly, studies from South Asia were retrieved using "exp Asia, Southern/" in MEDLINE and "exp South Asia/" in Embase. More complex queries were required for broader regions such as Europe excluding the UK and North America excluding the US, where multiple geographic terms were combined to ensure precise segmentation.

To assess temporal trends, publication searches were refined to include results from 2014 to 2024, using "(2014* or 2015* or 2016* or 2017* or 2018* or 2019* or 2020* or 2021* or 2022* or 2023* or 2024*).dt". The total number of EHR-related publications retrieved from MEDLINE and Embase was recorded separately for the .mp. search and the MeSH/Emtree search, allowing for an assessment of geographic representation over time.

For transparency and reproducibility, the full set of search queries, including all regional filters, is available on [GitHub](#), along with the number of publications retrieved for each geographic region and year. These results have been summarized in Tables S1.1–S1.4 to provide a structured overview of the search results.

To contextualize these findings, we used global population estimates from the 2024 Revision of World Population Prospects, published by the Population Division of the United Nations Department of Economic and Social Affairs (United Nations Department of Economic and Social Affairs Population Division, 2024). These estimates allowed for a direct comparison between research output and population size, offering a more equitable assessment of geographic disparities in EHR-related publications. For consistency across the study period, we assume that regional population proportions have remained relatively stable since 2014, allowing for meaningful comparisons between the 2024 population estimates and EHR publication proportions from 2014 onward. The original dataset and the R script used for data processing are publicly available on [GitHub](), ensuring full reproducibility. The processed dataset was then used to generate Figure 2 in the main document, and Figure S1 and **Error! Reference source not found.** in Section S1.2, which visualizes regional disparities in EHR research representation over the past decade.

The top panels of Figure S1 and **Error! Reference source not found.** illustrate the annual number of EHR-related publications indexed in MEDLINE and Embase from 2014 to 2024. In the .mp. keyword search, both databases exhibit fluctuations over time but show an overall upward trend, with Embase experiencing more pronounced growth. In contrast, in the MeSH/Emtree search, MEDLINE has seen a decline in EHR-related publications in recent years, while Embase has shown steady growth since 2017.

The middle and bottom panels of Figure S1 and **Error! Reference source not found.** depict the proportional representation of EHR research by geographic region, revealing persistent disparities. In Figure S1 (.mp. search), the United States share in MEDLINE declined from 57.8% in 2014 to 41.2% in 2024, while the United Kingdom's share fluctuated but ultimately increased from 7.5% to 10.0%. A similar trend was observed in Embase, where the US contribution dropped from 48.5% to 39.7%, while the UK's representation varied over time, ranging from 6.1% to 10.2% before settling at 8.3% in 2024. However, despite these shifts, the US and the UK together still accounted for more than half of all MEDLINE-indexed EHR studies in 2024, contributing 51.2% of publications, and remained dominant in Embase as well, where they collectively represented 48% of studies. This concentration of research output in just two high-income countries remains disproportionate to their combined population share of only 5.1% of the global total. Meanwhile, underrepresented regions, such as Africa (18.6% of the global population), South Asia (24.2%), and China (17.5%), have remained marginally represented in EHR research. In 2024, Africa contributed only 2.5% of EHR publications in both MEDLINE and Embase. Similarly, South Asia accounted for 1.4% of publications in MEDLINE and 1.8% in Embase, while China's representation stood at 6.5% in MEDLINE and 5.1% in Embase.

In **Error! Reference source not found.** (MeSH/Emtree search), similar patterns emerge, but the controlled vocabulary indexing results in slightly different proportions. In MEDLINE, the share of US-based studies declined from 58.9% in 2014 to 39.4% in 2024, while in Embase, the decrease was from 49.9% to 39.1%. The UK's contribution remained substantial in both databases, increasing from 7.8% to 11.5% in MEDLINE and fluctuating between 7.1% and 11.1% in Embase before settling at 10.5% in 2024. Together, the US and UK accounted for 50.9% of MEDLINE publications and 49.6% of Embase publications in 2024, reinforcing the continued dominance of high-income countries in

EHR research despite their small share of the global population. While African representation was slightly higher in MEDLINE, reaching 3.1% in 2024, its share in Embase declined slightly from 2.5% in the .mp. search to 2.3% in the Emtree search. Similarly, South Asia contributed just 1.1% of studies in MEDLINE and 1.7% in Embase, and China accounted for 4.1% in MEDLINE and 4.4% in Embase.

## S1.1.  Supplementary Tables

Table S1.1 Proportion of EHR Publications Over Time by Geographic Region using .mp. search (Medline)

| Year | 2014 | 2015 | 2016 | 2017 | 2018 | 2019 | 2020 | 2021 | 2022 | 2023 | 2024 | Population Proportion |
|---|---|---|---|---|---|---|---|---|---|---|---|---|
| **US** | 57.8% | 54.5% | 54.4% | 49.6% | 50.5% | 46.7% | 44.2% | 43.9% | 41.1% | 45.7% | 41.2% | 4.3% |
| **North America without US** | 4.2% | 4.1% | 3.6% | 4.9% | 3.9% | 4.3% | 4.5% | 5.2% | 5.0% | 5.0% | 2.9% | 2.1% |
| **UK** | 7.5% | 8.5% | 8.2% | 8.9% | 8.9% | 9.6% | 8.3% | 8.9% | 8.7% | 7.9% | 10.0% | 0.8% |
| **Europe without UK** | 15.6% | 17.4% | 16.0% | 14.7% | 14.8% | 13.4% | 13.8% | 13.4% | 14.0% | 11.2% | 17.6% | 8.5% |
| **China** | 0.9% | 1.7% | 1.7% | 2.9% | 4.1% | 5.0% | 8.2% | 6.6% | 5.8% | 5.6% | 6.5% | 17.5% |
| **East Asia without China** | 2.4% | 3.5% | 4.0% | 3.1% | 3.6% | 4.7% | 3.4% | 4.3% | 3.2% | 3.2% | 4.2% | 2.8% |
| **South Asia** | 0.2% | 0.4% | 0.7% | 1.4% | 0.7% | 1.3% | 1.4% | 1.2% | 2.3% | 2.4% | 1.4% | 24.2% |
| **Southeast Asia** | 0.3% | 0.1% | 0.9% | 0.9% | 1.3% | 1.4% | 1.7% | 1.2% | 1.6% | 1.4% | 1.3% | 8.5% |
| **Africa** | 1.4% | 1.3% | 1.2% | 2.0% | 1.9% | 1.9% | 2.2% | 2.6% | 3.0% | 2.5% | 2.5% | 18.6% |
| **Middle East** | 2.4% | 1.6% | 1.9% | 3.3% | 2.9% | 2.9% | 4.0% | 4.4% | 4.2% | 4.1% | 4.2% | 4.6% |
| **Aus-NZ** | 3.9% | 3.1% | 3.2% | 4.3% | 3.0% | 4.7% | 3.1% | 4.0% | 6.9% | 6.4% | 4.5% | 0.4% |
| **Multiple & Other** | 3.3% | 3.7% | 4.4% | 4.1% | 4.4% | 4.0% | 5.1% | 4.2% | 4.4% | 4.6% | 3.5% | 7.7% |
| **Total Number** | 1613 | 1737 | 1659 | 1602 | 1714 | 1909 | 2201 | 1889 | 1371 | 1306 | 2489 | |

Table S1.2 Proportion of EHR Publications Over Time by Geographic Region using .mp. search (Embase)

| Year | 2014 | 2015 | 2016 | 2017 | 2018 | 2019 | 2020 | 2021 | 2022 | 2023 | 2024 | Population Proportion |
|---|---|---|---|---|---|---|---|---|---|---|---|---|
| US | 48.5% | 46.0% | 48.3% | 45.5% | 46.9% | 46.6% | 44.6% | 45.2% | 42.8% | 43.1% | 39.7% | 4.3% |
| North America without US | 4.5% | 5.0% | 4.4% | 4.8% | 4.7% | 4.6% | 4.6% | 4.4% | 3.8% | 3.4% | 3.9% | 2.1% |
| UK | 10.2% | 10.0% | 8.2% | 8.0% | 6.5% | 6.9% | 6.1% | 7.3% | 7.6% | 8.9% | 8.3% | 0.8% |
| Europe without UK | 15.2% | 15.1% | 14.6% | 14.4% | 12.9% | 12.2% | 12.1% | 11.8% | 12.8% | 11.9% | 14.6% | 8.5% |
| China | 1.7% | 1.6% | 2.1% | 2.5% | 3.2% | 3.9% | 6.9% | 5.5% | 5.1% | 5.0% | 5.1% | 17.5% |
| East Asia without China | 2.9% | 4.2% | 4.5% | 5.4% | 4.3% | 5.1% | 3.7% | 4.1% | 4.7% | 4.5% | 5.3% | 2.8% |
| South Asia | 0.6% | 0.9% | 1.4% | 1.8% | 1.6% | 1.7% | 2.2% | 2.3% | 2.7% | 2.2% | 1.8% | 24.2% |
| Southeast Asia | 1.1% | 1.1% | 1.2% | 1.6% | 2.2% | 1.9% | 2.1% | 1.9% | 2.1% | 2.3% | 1.9% | 8.5% |
| Africa | 1.3% | 1.6% | 2.2% | 2.8% | 2.0% | 2.4% | 2.4% | 2.0% | 2.2% | 2.3% | 2.5% | 18.6% |
| Middle East | 2.4% | 1.9% | 2.3% | 3.5% | 3.1% | 3.5% | 4.1% | 5.0% | 4.7% | 5.2% | 4.8% | 4.6% |
| Aus-NZ | 3.5% | 3.3% | 3.1% | 2.7% | 3.4% | 3.4% | 3.5% | 2.7% | 4.5% | 3.9% | 3.5% | 0.4% |
| Multiple & Other | 8.1% | 9.4% | 7.6% | 7.0% | 9.1% | 7.8% | 7.6% | 7.8% | 7.0% | 7.4% | 8.6% | 7.7% |
| Total Number | 2469 | 2635 | 2404 | 2389 | 2788 | 3257 | 3845 | 4151 | 4577 | 4650 | 4564 | |

Table S1.3 Proportion of EHR Publications Over Time by Geographic Region using MeSH search (Medline)

| Year | 2014 | 2015 | 2016 | 2017 | 2018 | 2019 | 2020 | 2021 | 2022 | 2023 | 2024 | Population Proportion |
|---|---|---|---|---|---|---|---|---|---|---|---|---|
| US | 58.9% | 52.8% | 55.1% | 50.0% | 51.1% | 44.5% | 46.3% | 43.4% | 38.0% | 41.0% | 39.4% | 4.3% |
| North America without US | 4.1% | 4.7% | 3.9% | 5.3% | 4.0% | 5.2% | 5.3% | 6.1% | 5.0% | 5.2% | 3.2% | 2.1% |
| UK | 7.8% | 8.2% | 7.3% | 10.3% | 10.9% | 11.4% | 10.3% | 10.6% | 11.7% | 8.7% | 11.5% | 0.8% |
| Europe without UK | 15.6% | 17.6% | 15.7% | 15.8% | 15.0% | 14.0% | 12.9% | 14.1% | 15.5% | 15.8% | 23.0% | 8.5% |
| China | 0.7% | 1.7% | 1.7% | 2.7% | 3.6% | 5.4% | 5.5% | 4.4% | 5.8% | 5.5% | 4.1% | 17.5% |
| East Asia without China | 1.9% | 3.4% | 3.9% | 2.8% | 3.0% | 3.4% | 3.7% | 5.1% | 4.1% | 4.2% | 3.9% | 2.8% |
| South Asia | 0.2% | 0.4% | 0.6% | 1.2% | 0.7% | 1.3% | 1.7% | 1.6% | 3.2% | 2.6% | 1.1% | 24.2% |
| Southeast Asia | 0.4% | 0.2% | 0.9% | 0.5% | 0.7% | 1.2% | 0.8% | 1.0% | 0.9% | 0.6% | 1.1% | 8.5% |
| Africa | 1.4% | 1.7% | 1.1% | 1.6% | 1.7% | 1.8% | 3.1% | 3.3% | 3.5% | 4.2% | 3.1% | 18.6% |
| Middle East | 2.2% | 1.4% | 1.8% | 2.9% | 2.1% | 2.8% | 3.2% | 2.5% | 3.8% | 4.2% | 2.2% | 4.6% |
| Aus-NZ | 4.2% | 3.4% | 3.5% | 4.2% | 2.5% | 5.3% | 2.7% | 5.1% | 6.7% | 6.1% | 4.5% | 0.4% |
| Multiple & Other | 2.8% | 4.5% | 4.5% | 2.8% | 4.8% | 3.8% | 4.5% | 2.9% | 1.8% | 1.9% | 2.9% | 7.7% |
| Total Number | 1182 | 1277 | 1077 | 932 | 975 | 1006 | 904 | 611 | 342 | 310 | 918 | |

Table S1.4 Proportion of EHR Publications Over Time by Geographic Region using Emtree search (Embase)

| Year | 2014 | 2015 | 2016 | 2017 | 2018 | 2019 | 2020 | 2021 | 2022 | 2023 | 2024 | Population Proportion |
|---|---|---|---|---|---|---|---|---|---|---|---|---|
| **US** | 49.9% | 47.7% | 48.3% | 44.9% | 47.0% | 45.8% | 44.5% | 45.1% | 42.0% | 42.5% | 39.1% | 4.3% |
| **North America without US** | 4.4% | 5.5% | 4.7% | 4.7% | 4.5% | 4.4% | 4.7% | 4.0% | 3.7% | 3.3% | 3.8% | 2.1% |
| **UK** | 10.2% | 9.8% | 9.9% | 10.7% | 9.0% | 9.8% | 7.7% | 9.8% | 9.9% | 11.1% | 10.5% | 0.8% |
| **Europe without UK** | 14.8% | 14.1% | 13.9% | 14.0% | 12.4% | 11.6% | 11.7% | 11.7% | 13.6% | 12.3% | 15.1% | 8.5% |
| **China** | 1.6% | 1.6% | 2.3% | 2.4% | 3.2% | 3.7% | 6.4% | 4.9% | 4.4% | 4.6% | 4.4% | 17.5% |
| **East Asia without China** | 2.2% | 3.4% | 3.7% | 4.5% | 3.8% | 4.0% | 3.1% | 3.9% | 4.2% | 4.2% | 4.8% | 2.8% |
| **South Asia** | 0.7% | 0.9% | 1.4% | 1.6% | 1.3% | 1.6% | 2.1% | 2.1% | 2.5% | 1.9% | 1.7% | 24.2% |
| **Southeast Asia** | 1.1% | 1.2% | 1.3% | 1.6% | 2.4% | 1.7% | 2.2% | 1.8% | 2.2% | 2.3% | 1.8% | 8.5% |
| **Africa** | 1.3% | 1.7% | 1.9% | 2.9% | 1.9% | 2.1% | 2.2% | 1.9% | 2.0% | 2.2% | 2.3% | 18.6% |
| **Middle East** | 2.4% | 1.7% | 2.3% | 3.4% | 2.9% | 3.3% | 4.0% | 4.8% | 4.5% | 5.0% | 4.3% | 4.6% |
| **Aus-NZ** | 3.8% | 3.4% | 2.7% | 2.8% | 3.3% | 3.7% | 3.5% | 2.6% | 4.3% | 3.7% | 3.3% | 0.4% |
| **Multiple & Other** | 7.6% | 9.1% | 7.5% | 6.4% | 8.2% | 8.2% | 8.0% | 7.4% | 6.7% | 6.9% | 8.9% | 7.7% |
| **Total Number** | 2147 | 2235 | 1980 | 1887 | 2184 | 2651 | 3116 | 3420 | 3854 | 3987 | 4028 | |

## S1.2. Supplementary Figures

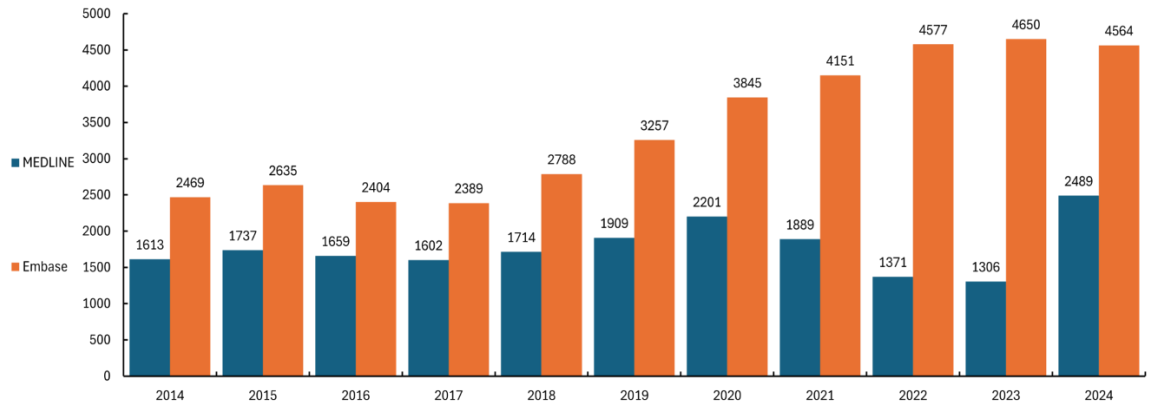

Figure S1a Total Number of EHR Publications Over Time (.mp.)

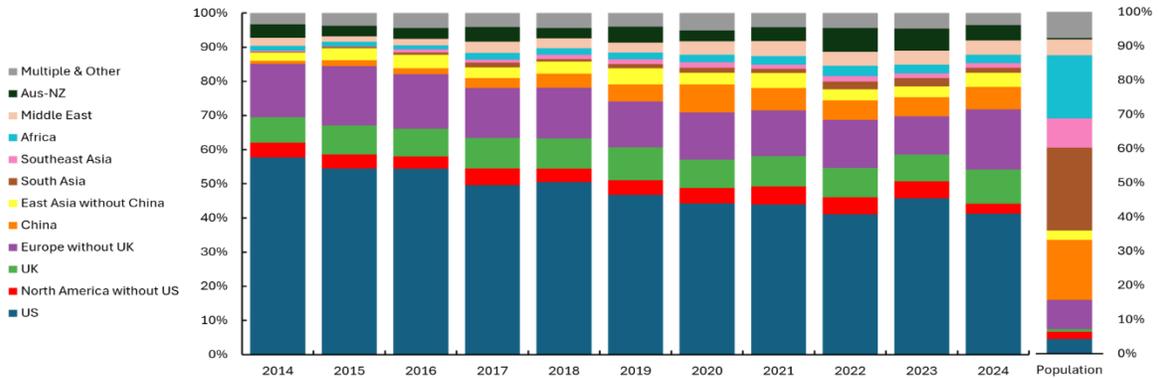

Figure S1b Proportion of EHR Publications Over Time by Study Population in Medline (.mp.)

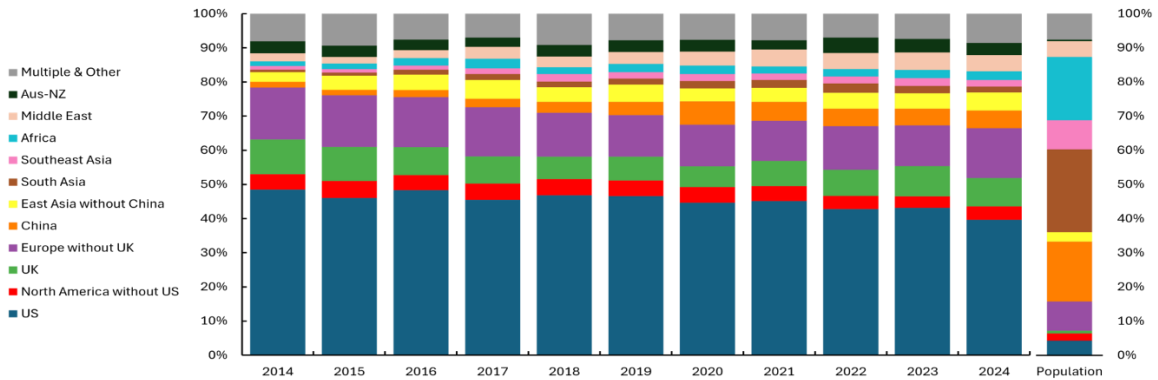

Figure S1c Proportion of EHR Publications Over Time by Study Population in Embase (.mp.)

Figure S1 Trends in EHR-related research indexed in Medline and Embase from 2014 to 2024, using the multi-purpose (.mp.) keyword search. Figure S1a: Annual number of EHR-related publications identified through the keyword-based (.mp.) search. Figure S1b: Proportional representation of EHR publications in Medline by study population region, based on the keyword search. Figure S1c: Proportional representation of EHR publications in Embase by study population region, based on the keyword search. In Figures S1b and S1c, the rightmost stacked bar represents the global population distribution estimated from the 2024 Revision of World Population Prospects, published by the Population Division of the United Nations Department of Economic and Social Affairs (United Nations Department of Economic and Social Affairs Population Division, 2024).

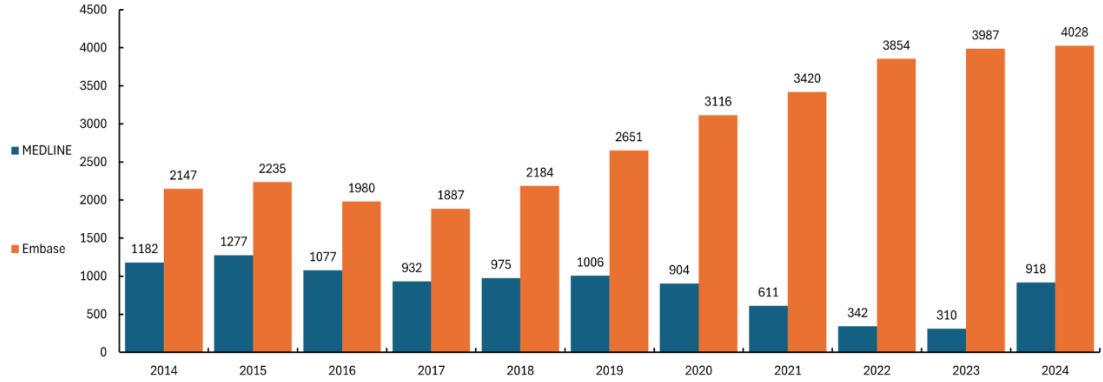

Figure S2a Total Number of EHR Publications Over Time (MeSH/Emtree)

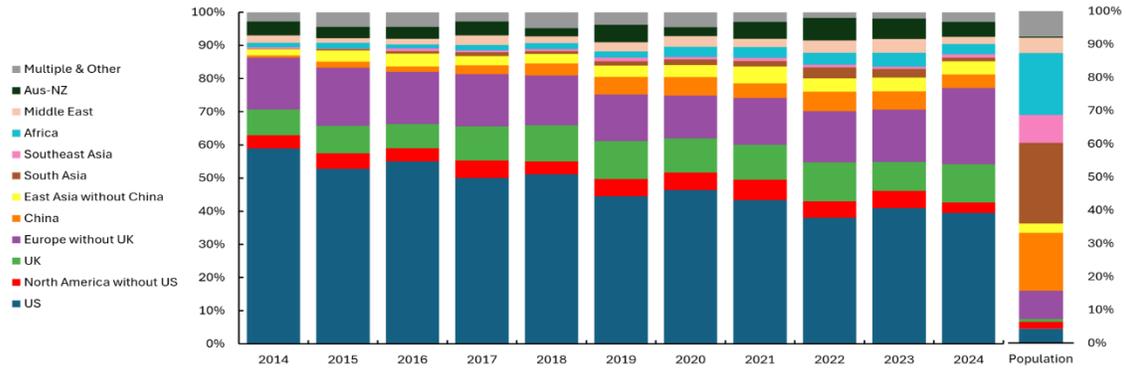

Figure S2b Proportion of EHR Publications Over Time by Study Population in MEDLINE (MeSH)

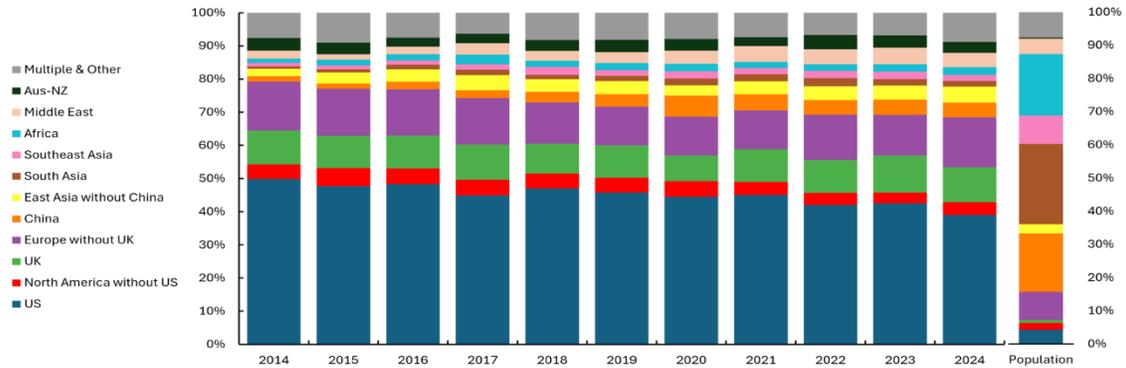

Figure S2c Proportion of EHR Publications Over Time by Study Population in Embase (Emtree)

Figure S2 Trends in EHR-related research indexed in Medline and Embase from 2014 to 2024, using the MeSH and Emtree heading search, respectively. Figure S2a: Annual number of EHR-related publications identified through the MeSH/Emtree search. Figure S2b: Proportional representation of EHR publications in Medline by study population region, based on the MeSH search. Figure S2c: Proportional representation of EHR publications in Embase by study population region, based on the Emtree search. In Figures S2b and S2c, the rightmost stacked bar represents the global population distribution estimated from the 2024 Revision of World Population Prospects, published by the Population Division of the United Nations Department of Economic and Social Affairs (United Nations Department of Economic and Social Affairs Population Division, 2024).

**S2. Supporting Information for GWAS Representation Analysis in Chapter 2.2**

This section details the data acquisition, preprocessing, and visualization methods used to analyze disparities in Genome-Wide Association Studies (GWAS) research representation in Chapter 2.2 of the main document. The methodology ensures reproducibility, and the complete code is available on [GitHub](GitHub).

### S2.1. Data Acquisition and Preprocessing

The dataset for this analysis was obtained from the NHGRI-EBI Catalog of Human Genome-Wide Association Studies (GWAS Catalog, available at https://www.ebi.ac.uk/gwas/home), using the latest available ancestry dataset as of May 13, 2025. The original dataset was provided as a tab-separated values (TSV) file, containing 198,017 observations across 12 column features. To facilitate further analysis, the dataset was preprocessed using R (R Core Team, 2025).

To ensure consistency, GWAS entries were ordered chronologically. Ancestral classifications were standardized to address inconsistencies in labeling across studies, consolidating populations into ten broad categories: European, East Asian, South Asian/Other Asian, African, Hispanic/Latino, Greater Middle Eastern, Oceanic, Other, Multiple, and Not Reported. Entries labeled as "Not Reported" were excluded from further analysis, while those with mixed ancestry were assigned to the "Multiple" category.

The reclassification process involved grouping populations with similar genetic and geographic backgrounds. African ancestry encompassed Sub-Saharan African, African American, and Afro-Caribbean populations. Asian ancestry was divided into East Asian and South Asian/Other Asian groups, reflecting distinctions in genetic studies. Greater Middle Eastern ancestry included individuals of Middle Eastern, North African, or Persian descent. This restructuring facilitated a more consistent and equitable comparison across studies.

For the records with the four highlighted PubMed IDs mentioned in the main text, we conducted the following actions to address the overlapping samples. For the studies with PubMed IDs 30104761 and 34662886, the primary source of overlap occurred within the control groups, which were reused across multiple analyses. To address this, we extracted control group sizes from sample descriptions and identified the largest reported control group size for each study. The actual contribution of these studies was then calculated as the sum of all case group sizes plus the maximum control group size, ensuring that each individual was only counted once.

For the study with PubMed ID 32888493, case and control groups were not explicitly specified in the descriptions. Instead, each row reported the total sample size, with both individual ancestry-level counts and combined totals across all ancestry groups. To prevent double counting, we extracted the maximum sample size for each ancestry category and removed the combined category rows. The final contribution of this study was determined by summing the maximum reported sample size for each ancestry group.

For the study with PubMed IDs 34594039 and 39024449, descriptions varied between case/control-specific counts and total sample sizes. Given that total sample size descriptions were generally larger than the sum of corresponding case and control groups, we prioritized using rows without case/control specification. To ensure consistency, we retained only the maximum sample size for each ancestry category, representing the study's actual contribution.

### S2.2. Visualization of GWAS Representation Disparities

To illustrate disparities in GWAS representation across different ancestry groups, visualizations were generated using R package ggplot2 (Wickham, 2016). A time series dataset was created by extracting unique dates from the DATE column. Cumulative participant counts were calculated for each ancestry category and visualized as stacked area plots in the upper left section of Figure 3 in the main document, capturing the evolution of representation over time.

In addition to cumulative representation, the proportional representation of each ancestry group was calculated relative to the total GWAS sample size at each time point and visualized in the lower left section of Figure 3.

### S2.3.     Comparison with Global Population Distributions

To provide context for GWAS representation disparities, 2025 global population estimates were sourced from Worldometer (Worldometer, 2025). Since the Worldometer classification system did not directly align with the GWAS dataset, a manual mapping process was conducted to aggregate populations into broad ancestry categories. Population estimates for each ancestry group were derived by summing the populations of countries where that ancestry is predominant. For instance, the African ancestry group was estimated using the total population of African countries, excluding individuals of African ancestry residing in non-African countries, such as the United States or the United Kingdom. Consequently, these estimates serve as an approximate reference rather than a precise representation of global ancestry distribution.

East Asian ancestry was derived from the Eastern Asia subregion, while South Asian/Other Asian ancestry included populations from Southern Asia and South-Eastern Asia. European ancestry was mapped from the total population of Europe. Greater Middle

Eastern ancestry encompassed populations from Western Asia and Northern Africa region. African ancestry was represented by the total population of Eastern Africa, Western Africa, Middle Africa, and Southern Africa. Hispanic/Latino ancestry was approximated using the total population of Latin America and the Caribbean. Oceanic ancestry was based on the total population of Oceania. These aggregated estimates, shown in the upper right section of Figure 3, provide a baseline reference.

### S2.4. Supplementary Tables

Table S2.1 Proportion of GWAS study representation as of April 25, 2025, alongside estimated global population proportions for each ancestry category in 2025. GWAS proportions indicate the percentage of participants in genome-wide association studies, while population proportions are based on Worldometer data (Worldometer, 2025). The "Multiple" category does not have a direct global population estimate.

| Ancestry Category | GWAS Proportion (%) | Global Population Proportion (%) |
|---|---|---|
| European | 87.0 | 9.13 |
| East Asian | 5.13 | 20.07 |
| South Asian/Other Asian | 0.42 | 33.84 |
| African | 1.05 | 15.47 |
| Hispanic/Latino | 0.83 | 8.11 |
| Greater Middle Eastern | 0.02 | 7.17 |
| Oceanic | 0.01 | 0.57 |
| Other | 0.04 | 5.64 |
| Multiple | 0.53 | |